\documentclass[prb,twocolumn,showpacs]{revtex4}

\usepackage{amssymb,amsmath,bm,epsfig,graphics,subfigure}

\begin{document}

\title{Nodal Surfaces in Photoemission from Twisted Bilayer Graphene}
\author{Anshuman Pal and E. J. Mele}
    \email{mele@physics.upenn.edu}
    \affiliation{Department of Physics and Astronomy University of Pennsylvania Philadelphia PA 19104  \\}
\date{\today}

\begin{abstract}
Selection rules and interference effects in angle resolved photoemission spectra from twisted graphene bilayers are studied
within a long wavelength theory for the electronic structure. Using a generic model for the interlayer coupling, we identify
features in the calculated ARPES momentum distributions that are controlled by the singularities and topological character of
its long wavelength spectrum. We distinguish spectral features that are controlled by single-layer singularities in the
spectrum, their modification by gauge potentials in each layer generated by the interlayer coupling, and new energy-dependent
interference effects that directly probe the interlayer coherence. The results demonstrate how the energy- and polarization-
dependence of ARPES spectra can be used to characterize the interlayer coupling in twisted bilayer graphenes.
\end{abstract}

 \maketitle

\medskip
\section{Introduction}
Angle resolved photoemission spectroscopy (ARPES) is a well developed tool for mapping out electronic bands in solids. Recent
applications to single layer and few-layer graphenes have demonstrated that, in addition to obtaining the energies and momenta
of the quasiparticle bands with high resolution, the measured variations of the photoemission intensities also encode
information about the phase structure of the wavefunctions \cite{Shirley,Ohta,Rotenberg,Kern,Louie}. The matrix elements that
couple the initial quasiparticle states to the outgoing free electron states are momentum dependent and produce modulations in
the intensities of ARPES momentum distributions at a constant initial state energy. These matrix elements typically contain a
coherent superposition of amplitudes for emission from the different sites and layers that are represented in the initial state
wavefunctions. These in turn depend on the (conserved) value of the crystal momentum parallel to the surface plane as well as
the energy and polarization state of the exciting radiation.

\medskip
Interpreting these intensity modulations presents a complex problem that has been addressed recently for the case of single
layer and some forms of bilayer graphene \cite{Rotenberg,Louie}. Graphene presents a nearly ideal platform for exploring this
phenomena for two reasons: (i) Graphene is a two dimensional material in which there is no dispersion of the quasiparticle
bands due to a third (perpendicular) component of the crystal momentum. (ii) The quasiparticle band structure contains {\it
point singularities} around which the internal phases in its wavefunctions undergo a complete twist.  Indeed, interfering
amplitudes in ARPES from single layer graphene have been observed and have striking consequences. The photoemission intensity
on a constant energy momentum space contour {\it vanishes} along a particular direction (labelled the ``dark corridor"
\cite{Kern}) where the emission amplitudes from the two sublattices turns out to exactly cancel. The orientations of these dark
corridors are rotated for photoemission from states near symmetry-related zone corner points when resolved in the extended
Brillouin zone. For $AB$-stacked bilayer graphene the emission patterns are more complex but they have been analyzed similarly
to identify the sign of the dominant interlayer Hamiltonian matrix elements between neighboring sites in the adjacent layers
\cite{Louie}. Theory predicts that single layer graphene with a strong Rashba spin-orbit interactions should have a distinct
photoemission signature in its {\it spin}-resolved ARPES reflecting the entanglement of its spin and orbital (pseudospin)
degrees of freedom \cite{FKIRashba}.

\medskip
All of these previous theoretical analyses take as their starting points a (presumed known) model for the low energy electronic
Hamiltonian and deduce the consequences for the ARPES momentum distributions as a function of the initial state energy. In this
paper we take the complementary point of view and consider the ARPES signatures of a {\it generic} Hamiltonian for a graphene
bilayer. Our approach is motivated by our interest in better understanding the electronic structure of ``twisted" multilayer
graphenes where the symmetry axes of the two layers are rotationally misaligned \cite{Berger,Hass2}. It is now widely
appreciated that this misalignment has a profound effect on the electronic coherence between neighboring layers
\cite{LpD,Latil,Shallcross,GTdL,MeleRC,Bistritzer}, though developing a microscopic theory for it has proven to be an elusive
goal. Current theories indicate that this problem is very rich since kinematical constraints  define different regions of
energy and momentum where the two layers can act as ``strongly coupled" or ``nearly decoupled" even for a single structure.
These energy-momentum sectors depend on the misalignment angle in a way that is not yet fully understood. For small angles of
misalignment theory predicts that new spectral features and narrow bands emerge at low energy \cite{Bistritzer,Guinea}.
Existing experiments are providing conflicting information about the nature of the interlayer coupling
\cite{LiAndrei,Sprinkle,GeimAndrei,STM,Hicks} a difficulty that may arise from physical differences between twisted graphenes
that are made by different experimental methods \cite{MelePRB}.

\medskip
Since both intra-layer and interlayer  electronic coherence in multilayer graphenes can be identified by their interference
signatures in ARPES, we suggest that deducing the coupling from experiment on twisted graphenes rather than attempting to
calculate it from  microscopic theory is a promising direction. In this paper we follow this approach and present calculations
of  ARPES intensities using a generic model for a twisted graphene bilayer. The inputs to the model are the Dirac models for
two decoupled layers, a momentum offset due to the rotational misalignment, an electrostatic asymmetry (bias) between the
layers and a {\it general} long wavelength matrix that couples the pseudospin amplitudes on neighboring layers. From this
starting point we calculate the electron momentum distributions at fixed energy and the ARPES distributions which are weighted
by the momentum dependent photoemission probabilities for various polarization states of the exciting radiation. A careful
analysis of the ARPES momentum distributions shows that the ``dark corridor" known for single layer graphene changes its shape
for twisted graphene and allows one to discriminate between various models for the coupling between layers. We present
calculations of the ARPES intensities that illustrate this effect and a symmetry analysis of the relevant matrix that allows us
to interpret the spectra. The results demonstrate how experimental study of the ARPES spectra as a function of the energy and
polarization of the exciting radiation can be used to fully characterize the interlayer coherence in these multilayer
structures.

\medskip
In this paper we first briefly review the long wavelength description of twisted bilayer graphene with a theory that is
appropriate to small rotation angles and in Section II we derive an expression for the photoemission transition matrix elements
based on this model. In section III we survey the topological transitions in the band structure that occur in this model as the
Hamiltonian parameters are varied. Section IV shows the calculated ARPES momentum distributions showing a rich assortement of
interference phenomena which are unique to twisted multilayer graphenes. Section V gives an analysis of the interference
effects with in terms of the underlying symmetries of the model. We present an analysis of the interference patterns in terms
of its ``energy-flattened" nodal surfaces in momentum space which provides a useful diagnostic for various forms of interlayer
coherence. A further discussion of the experimental signatures of these predictions is given in Section VI.

\section{Long Wavelength Model}

\subsection{Small Angle Twisted Bilayers}
Our work uses a long wavelength low energy theory of the electronic states in graphene bilayers. In single layer graphene the
electrons propagate on a honeycomb lattice with two sublattice sites labelled $A$ and $B$. At low energies there are two
electronic bands that touch at $E=0$ at discrete points located at the Brillouin zone corners  $\nu {\bm K}$ where $\nu =
1(-1)$ denote the ${\bm K}({\bm K'})$ points. These bands disperse linearly around the contact points and are described by a
Hamiltonian at crystal momentum ${\bm k}$ that can be linearized in the difference ${\bm q} = {\bm k} - \nu {\bm K}$
\begin{eqnarray}\label{DiracS}
{\cal H}_{\rm S} = \hbar v_F {\bm \sigma} \cdot ( \nu q_x \hat e_x + q_y \hat e_y)
\end{eqnarray}
where ${\bm \sigma}$ are Pauli matrices acting in the space of sublattice (pseudospin) degree of freedom.

\medskip
For a graphene bilayer we adopt a four-component basis with a conventional ordering of its four site-layer degrees of freedom
$(A_1, \, B_1, \, A_2, \, B_2)$. When the symmetry axes of the two layers are aligned and the interlayer coupling is set to
zero the Hamiltonian for the two layers is simply a doubled version of the single layer Hamiltonian and can be written in the
form
\begin{eqnarray}
{\cal H}_{\rm B} = {\cal H}_{\rm S} \, \tau_0
\end{eqnarray}
where ${\bm \tau}$ are Pauli matrices acting on the layer degree of freedom. In $AB$ stacked graphene the symmetry axes of the
two layers are aligned and the two layers are coupled through a hybridization of amplitudes on different sublattice sites in
the two layers e.g. $(A_1, \, B_2)$ as described by the bilayer Hamiltonian
\begin{eqnarray}
{\cal H}_{\rm Bernal} = {\cal H}_{\rm S} \, \tau_0 + \frac{\gamma_1}{2} \left( \sigma_x \tau_x - \sigma_y \tau_y \right)
\end{eqnarray}
where $\gamma_1$ is the strength of the interlayer coupling.

\medskip
In this work we are concerned with the situation where the symmetry axes of the two layers are rotated by a relative angle
$\theta$. It is convenient to regard each of the layers as rotated by angles $\pm \theta/2$ with respect to a common reference
structure in which the ${\bm K}$ point is oriented along the $x$ direction as shown in Figure 1. With this convention the ${\bm
K}$ points in the two layers shift by momenta $\Delta {\bm K}_\pm$ which are, in complex vector notation
\begin{eqnarray}
\Delta {\bm K}_\pm = \left( e^{\pm i \theta/2} -1 \right) {\bm K}
\end{eqnarray}
The components of ${\bm q}$ are resolved in the original unrotated frame so that the effect of the rotation is to induce both a
shift of origin {\it and} the phases of the complex momenta appearing in ${\cal H}_{\rm B}$ according to the replacement rule

\begin{figure}
\begin{center}
  \includegraphics[angle=0,width=0.8\columnwidth]{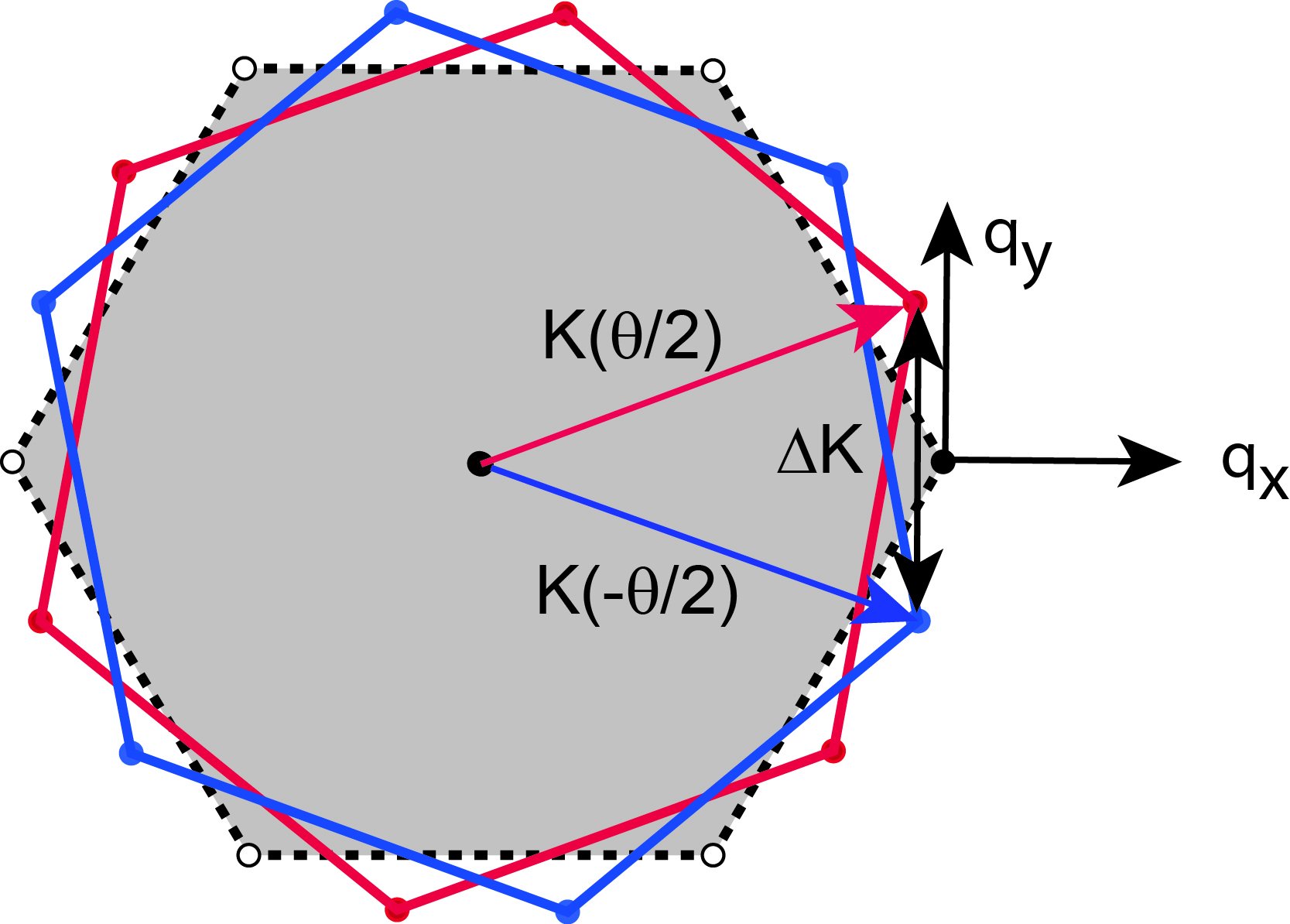}
  \caption{\label{Introfig} The Brillouin zones for each layer (red and blue denote different layers) of a twisted bilayer are rotated by angles $\pm \theta/2$
  with respect to a reference zone (dashed) with a  corner ${\bm K}$ aligned with the $x$ axis. In the long wavelength model presented in
  the text all momenta ${\bm q}$ are measured with respect to the reference Brillouin zone corner and resolved along the $(q_x,q_y)$ axes as shown.}
\end{center}
\end{figure}

\begin{eqnarray}
q_x  +i q_y &\mapsto& q_{x,\pm}' + i q_{y,\pm}' \nonumber\\
   q_{x,\pm}' + i q_{y,\pm}' &=& \left( q_x +i q_y - \left( e^{\pm i \theta/2} -1 \right) {\bm K} \right) e^{\mp
i \theta/2} \nonumber\\
\end{eqnarray}
Expressed in a common frame, the offset single-layer Dirac operators of Eqn. (\ref{DiracS}) are
\begin{eqnarray}\label{Diracpm}
{\cal H}_{\pm} = \hbar v_F {\bm \sigma} \cdot ( \nu q_{x,\pm}' \hat e_x + q_{y,\pm}' \hat e_y)
\end{eqnarray}
When the rotation angle $\theta$ is small the interlayer coupling across a twisted bilayer admits a description analogous to
the $\gamma$ term in Eqn. 3. To see this, before passing to the long wavelength limit (Eqn. 1) we partition the bilayer lattice
Hamiltonian ${\cal H}_{\rm lattice}$ into $2 \times 2$ layer diagonal and layer off diagonal blocks
\begin{eqnarray}\label{Hlatt}
{\cal H}_{\rm lattice} = \left(%
\begin{array}{cc}
  {\cal H}_1& {\cal T}(\vec r) \\
  {\cal T}^{\rm T} (\vec r)& {\cal H}_2 \\
\end{array}%
\right)
\end{eqnarray}
When $\theta$ is small the registry between layers is modulated in a supercell evolving smoothly through local zones with
$(A_1B_2)$ stacking, $(B_1A_2)$ stacking and $(A_1A_2)$ stacking. In this situation ${\cal T}(\vec r)$ in Eqn. (\ref{Hlatt}) is
a smooth, periodic and local $2 \times 2$ matrix function of $\vec r$ acting on the pseudospins in each layer. The smoothest
periodic matrix-valued function satisfying these constraints is
\begin{eqnarray}
{\cal T}(\vec r) = \hat t_0 + \sum_{n=1}^6 \,  \hat t_n \, e^{i \vec {\cal G}_n \cdot \vec r}
\end{eqnarray}
where $\hat t_n \, (n=0,6)$ are $2 \times 2$ matrix-valued constants and $\vec {\cal G}_n \, (n=1,6)$ are vectors in the first
star of superlattice reciprocal lattice vectors. The matrix coefficients $\hat t_n$ can be further simplified by exploiting
lattice and rotational symmetries. In a periodic supercell with its $AB$, $BA$ and $AA$ zones centered at positions $\vec
r_\alpha$, $\vec r_\beta$ and $\vec r_\gamma$ respectively, the coefficients $\hat t_n$ for the three even elements of $\vec
{\cal G}_n$ are
\begin{eqnarray}
\hat t_{n \, {\rm even}} = t_{\cal G} \left(%
\begin{array}{cc}
  e^{-i \vec G_n \cdot \vec r_\gamma} & e^{-i \vec G_n \cdot \vec r_\alpha}  \\
  e^{-i \vec G_n \cdot \vec r_\beta}  & e^{-i \vec G_n \cdot \vec r_\gamma}  \\
\end{array}%
\right) = t_{\cal G} \left(%
\begin{array}{cc}
  z & 1  \\
  z^*  & z  \\
\end{array}%
\right)
\end{eqnarray}
where $z=\exp(2 \pi i/3)$ and $\hat t_{n \, {\rm odd}} = \hat t_{n \, {\rm even}}^*$. Sublattice symmetry requires that the
constant matrix $\hat t_0$ has the form
\begin{eqnarray}
\hat t_0 = \left(%
\begin{array}{cc}
  c_{AA} & c_{AB} \\
  c_{BA} & c_{BB} \\
\end{array}%
\right)
\end{eqnarray}
In the small $\theta$ limit, the interlayer coupling is described by three real parameters $c_{AA}=c_{BB}$, $c_{AB} = c_{BA}$
and $t_{\cal G}$. One can choose these parameters by specifying the direct ($\gamma_1$) and skew ($\gamma_3, \,  \gamma_4$)
interlayer hopping parameters in the $AB$ registered zones of a twisted bilayer.  Table I makes this translation by giving the
constants in the modulated interlayer hopping model of a twisted bilayer in terms of the three interlayer intersite amplitudes.
Note that if the interlayer hopping amplitudes is long range, i.e. if it varies slowly on the scale of the graphene primitive
cell, then $t_{\cal G}$ is small compared to the constant term. When the skew hopping terms are isotropic ($\gamma_3=\gamma_4$)
the diagonal and off diagonal terms in the constant matrix have the same amplitude ($c_{AA} = c_{AB}$).

As examples, Table I evaluates these constants for two models. In Model I a direct interlayer term $\gamma_1$ is retained and
the skew hopping terms are arbitrarily set to zero. Even in this extremely short range model one finds that the constant terms
dominate the spatially modulated part of the interlayer Hamiltonian. In Model II the constants are fit to the
Slonczewski-Weiss-McClure (SWMc) parameterization of the interlayer amplitudes in Bernal graphite retaining its skew hopping
terms \cite{SWMc} and as expected the amplitude of the modulated term is even weaker. Note also that the SWMc parameterization
also shows a strong asymmetry between $\gamma_3$ and $\gamma_4$. These two amplitudes describe interlayer hopping processes at
the same range but in different crystallographic directions with respect to the graphene symmetry axes.  This asymmetry,
deduced from fitting experimental data, is not contained in  two-center tight binding theories that posit an isotropic
interlayer hopping model. The asymmetry arises from a crystal field anisotropy in the Wannier states appropriate to single
layer graphene.

\begin{table}
  \centering
\begin{tabular}{c c c c} 
\hline\hline 
Coefficient & Parameterization    & I & II  \\ [0.5ex] 
\hline 
$t_{\cal G}$ & $(\gamma_1 - \gamma_3)/9$ & 43.3  & 8.3  \\ 
$c_{AA}$ & $\gamma_4 + (\gamma_1 - \gamma_3)/3$ & 130.0 & 69.0  \\
$c_{AB}$ & $(\gamma_1 + 2 \gamma_3)/3$ & 130.0  & 340.0  \\ [1ex] 
\hline\hline 
\end{tabular}
\label{parameters} 
  \caption{Parameters (meV units) in interlayer hopping Hamiltonian for a twisted bilayer fitted to the
  interlayer intersite amplitudes $\gamma_1$, $\gamma_3$ and $\gamma_4$ in its Bernal stacked zones. Model I:
  $\gamma_1 = 390$ meV, $\gamma_3=\gamma_4=0$. Model II: $\gamma_1 = 390$ meV, $\gamma_3$ = 315 meV and $\gamma_4 = 44$ meV.}\label{SWMcC}
\end{table}

The constant matrix $\hat t_0$ is the supercell-average of the interlayer hopping operator \cite{MelePRB}. One can distinguish
three different kinds of behavior depending on whether its off diagonal elements are dominant, its diagonal elements are
dominant or they are equal. The SWMc parameterization realizes the first type of behavior since, as shown in Table
\ref{parameters}, it is dominated by its site off-diagonal parts: $c_{AB} \gg c_{AA}$.  Alternatively, any two center tight
binding model that assumes an isotropic interlayer hopping model is a member of the last family which describes a different
type of interlayer coherence as we show explicitly below.  Finally, as noted in our earlier work,  the absence of a significant
Fermi velocity renormalization in a family of twisted multilayer graphenes as well as some puzzling features in their measured
ARPES spectra can be understood using the the second class of interlayer models in which interlayer hopping on the {\it same}
sublattices controls the interlayer Hamiltonian. In this paper we will regard the constant matrix $\hat t_0$ as unknown a
priori and study the signatures of each of these three forms in its ARPES spectra.

In the four-band description of the twisted bilayer problem the interlayer coupling Hamiltonian ${\cal H}_{+,-} =(\Gamma_1
\sigma_x+ \Gamma_2 \sigma_0)\tau_x$ where $\Gamma_1 > \Gamma_2 $ in the first class, $\Gamma_2 > \Gamma_1$ in the second class
and $\Gamma_1 = \Gamma_2$ is a marginal state that separates them. Allowing for the possibility of different electrostatic
potentials $\pm V/2$ on the two layers we therefore consider the family of Hamiltonians in the $\nu=1$ valley
\begin{eqnarray}\label{HamT}
{\cal H}_{\rm T} &=&  \frac{\hbar v_F}{2} \left(   {\bm \sigma} \cdot {\bm q'_+}(\tau_0 + \tau_z)
 +   {\bm \sigma} \cdot {\bm q'_-}(\tau_0 -
\tau_z) \right) \nonumber\\
 &+& \frac{V}{2} \sigma_0 \tau_z + (\Gamma_1 \sigma_x+ \Gamma_2
\sigma_0)\tau_x
\end{eqnarray}
where the related Hamiltonian for $\nu=-1$ is obtained by the substitution ${\bm \sigma} \rightarrow -{\bm \sigma}^*$. The
Hamiltonian Eqn. (\ref{HamT}) is the starting point for the calculations we present in this paper. It is a function of the
rotation angle $\theta$,  the electrostatic potential difference $V$ and coupling coefficients $\Gamma_n$. Choices of
$\Gamma_n$ allow us to explore the space of symmetry allowed interlayer coupling models.

Note that the projection of the original lattice Hamiltonian ${\cal H}_{+,-}$ onto the Dirac basis changes the phases of the
interlayer terms because of the oscillation of the ${\bm {K(K')}}$ point pseudospin basis functions. Thus the interlayer in
(\ref{HamT}) are multiplied by phase factors $\exp[i(\vec G \cdot \vec d_i - \vec G' \cdot \vec d'_j)]$ where $\vec G (\vec
G')$ are either zero or primitive reciprocal lattice vectors in the two separate layers that connect equivalent zone corners
points and $\vec d_i (\vec d'_j)$ are the sublattice positions in the two layers. Simultaneous boosts by a triad of ($\vec G
,\vec G'$) pairs transforms the Hamiltonian (\ref{HamT}) among three pairs of zone corner points in which the twist-induced
offsets are $\Delta {\bm K}$ and its $\pm 2 \pi/3$-rotated counterparts. For definiteness in the remainder of this paper we
present calculations for the choice $\vec G = \vec G'=0$ as given in Eqn. (\ref{HamT}) describing emission from states near a
single ${\bm K}$ point.

\subsection{Photoemission Matrix Elements}
In angle resolved photoemission, an incident photon with energy $\hbar \omega$ excites a Bloch electron into an outgoing free
particle state whose energy $E_{\bm p} = p^2/2m_e$ and propagation direction ${\bm p} = ({\bm p_\|},p_z)$ are measured outside
the sample. Conservation of energy identifies the initial state energy $E_i$ with respect to the work function $W$
\begin{eqnarray}
E_i = E_{\bm p}+ W  - \hbar \omega  \nonumber
\end{eqnarray}
and conservation of the parallel component of momentum identifies the relative momentum ${\bm q}$ for emission near valley $\nu
{\bm K} + {\bm G}$
\begin{eqnarray}
{\bm q} = {\bm p_\|} - \nu {\bm K} - {\bm G} \nonumber
\end{eqnarray}

The incident light is coupled to the electrons through an interaction Hamiltonian
\begin{eqnarray}
{\cal H}_{\rm int} = -e  \vec {\bm v} \cdot \vec A_\omega  e^{-i \omega t}
\end{eqnarray}
where the velocity operators are obtained by differentiating Eqn. (\ref{HamT}):  $v_\mu = \partial {\cal H_{\rm T}}/\partial
q_{\mu}$ giving
\begin{eqnarray}\label{velocities}
\left(%
\begin{array}{c}
  v_x \\
  v_y \\
\end{array}%
\right) = \left(%
\begin{array}{cc}
  \cos \frac{\theta}{2} \, \tau_0 & \mp \sin \frac{ \theta}{2}\,  \tau_z \\
  \pm \sin \frac{ \theta}{2} \, \tau_z & \cos \frac{\theta}{2} \, \tau_0\\
\end{array}%
\right)
\left(%
\begin{array}{c}
  \sigma_x \\
  \sigma_y \\
\end{array}%
\right)
\end{eqnarray}
Equation \ref{velocities} defines two $4 \times 4$ matrix operators acting in the pseudospin-layer orbital space for each of
the two orthogonal polarizations. These operators are independent of momentum ${\bm q}$ but depend on the rotation angles $\pm
\theta/2$.

The matrix elements for photoexcitation with photon polarization $\hat e$ are
\begin{eqnarray}\label{M}
{\cal M}({\bm p}, {\bm q}) = \langle \psi^>_{\bm p} | \vec {\bm v} \cdot \hat e | \psi_{\bm q} \rangle
\end{eqnarray}
To evaluate Eqn. (\ref{M}) we project the outgoing plane wave state $\psi^>_{\bm p}$ onto the same basis used to represent the
initial state $\psi_{\bm q}$. Crucially, this projection introduces a phase difference $\phi = p_z d/\hbar$ between plane wave
amplitudes on the two layers separated by a vertical distance $d$. For photoemission near the graphene Brillouin zone corner,
by using excitation energies in the range 30-100 eV, $\phi$  can be varied over a range of approximately $4 \pi$. Note that the
{\it parallel} component of the momentum ${\bm p}_\|$ is conserved, laterally phase matching the outgoing state to the initial
Bloch state. Therefore, under typical experimental conditions the analogous lateral interference effects are small and nearly
energy independent. (They would be exactly zero for perpendicular emission.) These small effects are not included in our
calculations. Thus, written in the orbital-layer basis, the final state appearing in Eqn. (\ref{M}) in our model is
\begin{eqnarray}\label{pw}
\psi^>_{\bm p} = \frac{1}{2} \left(%
\begin{array}{c}
  1 \\
  1 \\
  e^{i \phi} \\
  e^{i \phi} \\
\end{array}%
\right) e^{i {\bm p} \cdot {\bm r}}
\end{eqnarray}
The matrix element in Eqn. ({\ref{M}) is the inner product of the operator Eqn. (\ref{velocities}) between an occupied
eigenstate of Eqn. (\ref{HamT}) and the plane wave final state given in  Eqn. (\ref{pw}).

Our analysis presents three different momentum distributions for interpretation of the ARPES intensities. The first is a map of
the spectral function at $A({\bm q},E_i)$ as a function of initial state energy $E_i$ unweighted by the transition matrix
elements
\begin{eqnarray}\label{Aofq}
A({\bm q},E_i) = \sum_n \,  \delta(E_n(\bm q) - E_i)
\end{eqnarray}
summed over occupied bands $n$ of Eqn. (\ref{HamT}). The second is a map of a momentum distribution of the simulated ARPES {\it
intensities} $I({\bm q},E_i)$ which includes the state dependent transition matrix elements
\begin{eqnarray}\label{Iofq}
I({\bm q},E_i) = \sum_n \,   |{\cal M}_n({\bm p}, {\bm q})|^2 \,  \delta(E_n(\bm q) - E_i)
\end{eqnarray}
As noted above the matrix elements $|{\cal M}({\bm p}, {\bm q})|^2$ also depend on the polarization state of the incident
light. The third is a map that ``flattens" the energy resolution of Eqn. (\ref{Iofq}), giving instead for each of the occupied
bands the distribution of just its momentum- and polarization-dependent squared matrix elements
\begin{eqnarray}\label{Mofq}
P_n({\bm q}) =   |{\cal M}_n({\bm p}, {\bm q})|^2
\end{eqnarray}
The summand in  Eqn. (\ref{Iofq}) is the product of the spectral functions (\ref{Aofq}) which expresses the kinematical
constraints on the experiment with the transition amplitudes (\ref{Mofq}) that encode the phase information.  Maps of the
unconstrained distributions $P_n({\bm q})$ are useful for exposing the quantum geometric origins of the interference patterns
that are accessible in the ARPES intensities. Experimentally one can determine the nodal structure of (\ref{Mofq}) by the
evolution of a series of spectra that sweep $E_i$ through the occupied bands.

\section{Topological Transitions}

The spectrum of the Hamiltonian in Eqn. (\ref{HamT}) exhibits topological transitions as a function of its interlayer coupling
parameters $\Gamma_1$ and $\Gamma_2$ as shown in Figure 2. In the literature on twisted graphenes a frequently used model
treats the case $\Gamma_1 = \Gamma_2$ \cite{LpD,Bistritzer,deGail} where the long wavelength coupling when the interlayer
amplitudes are {\it isotropic} functions of the relative position between two sites projected into the $xy$ plane. The spectrum
of this model shows a pair of Dirac singularities at $E=0$ and a saddle point singularity at higher energy where the two Dirac
cones merge and hybridize. A representative spectrum with these features is shown in the far left panel of Figure
\ref{bandsurfaceplotsI}. Perturbation theory in the interlayer coupling predicts a $\theta$-dependent renormalization of the
Dirac velocities of its low energy features which are largest in the limit of small rotation angles when the two Dirac features
are nearly congruent. Calculations beyond perturbation theory suggest that this low angle regime supports rich low energy
physics with the emergence of new low energy nearly flat bands whose narrow bandwidth oscillates as a function of the rotation
angle \cite{Bistritzer,Guinea}.

Symmetry does not require $\Gamma_1 = \Gamma_2$ and in earlier work \cite{MelePRB} we noted that empirical parameterization
schemes, most notably the Slonczewski-Weiss-McClure (SWMc) model suggest that there can be a very strong asymmetry between
these two parameters. This produces striking changes to both the symmetry and the topology of the electronic bands both at low
energy where the Dirac points are effectively decoupled and at higher energy where they merge.  Using the results of Table I,
we see that the conventional SWMc parameterization describes a situations where $\Gamma_1 \gg \Gamma_2$, i.e. the
supercell-averaged interlayer coupling is dominated by its sublattice off diagonal terms. An extreme version of this situation
one can examine the case $\Gamma_1 \neq 0, \Gamma_2=0$. If the coupling strength $\Gamma_1$ is weak this model shares many
features of the isotropic model, with a slightly lower symmetry as shown in the middle panel of Figure \ref{bandsurfaceplotsI}.
In our earlier work \cite{MelePRB} we found that after a gauge transformation this theory can be cast into a form where it
describes two momentum-offset Dirac cones with {\it opposite} helicity coupled by a scalar (i.e. sublattice diagonal)
interlayer coupling matrix. In that analysis one finds that the interlayer coupling generates a gauge potential in each layer
that displaces the two Dirac nodes towards each other as the rotation angle is decreased. Decreasing the rotation angle {\it
increases} the effective coupling strength $\Gamma_1$. Ultimately, at a critical rotation angle the nodes merge and annihilate
as shown in Figure \ref{bandsurfaceplotsII}. For still smaller angles two new point singularities emerge describing two {\it
new} Dirac modes of a strongly interlayer-coupled limit. It is important to appreciate that the momentum difference between the
Dirac nodes is not determined purely geometrically by the rotation angle, but instead it is generally changed by the strength
and symmetry of the coupling between the layers. Also note that in the crossover regime the system has a spectrum reminiscent
of $AB$ stacked graphene, albeit on a reduced energy scale. The presence of the singularities at $E=0$ is symmetry protected
and, as is the case in $AB$-stacked graphene, cannot be removed unless the layer symmetry is lifted.

\begin{widetext}

\begin{figure}
\begin{center}
  \includegraphics[angle=0,width=0.8\columnwidth]{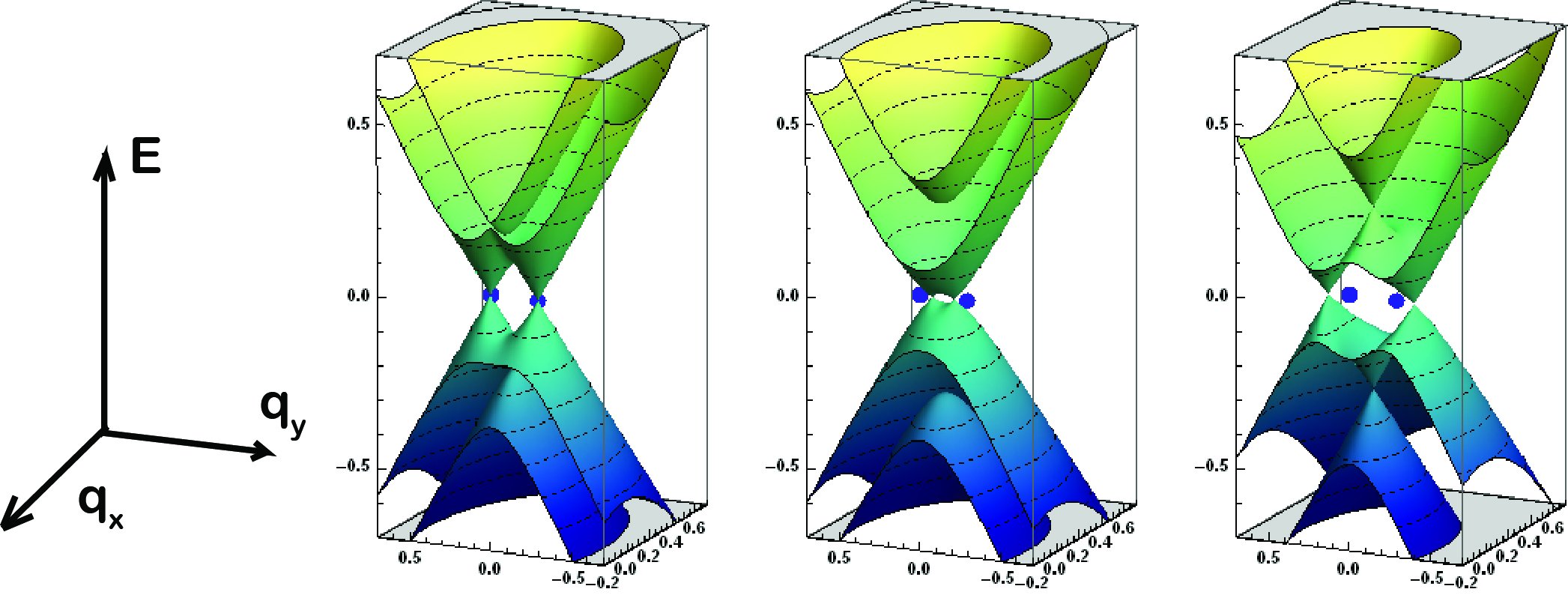}
  \caption{\label{bandsurfaceplotsI} Energy surfaces in the four band model for a twisted graphene bilayer for
  three different models for the long wavelength interlayer coupling matrix: (left panel) Model I with $\hat V_{\rm int} \propto (\mathbb{I} + \sigma_x)$,
  (middle panel) ($\Gamma_1 = \Gamma_2 = 0.05$) Model II with $\hat V_{\rm int} \propto \sigma_x$ in the weak coupling regime
  ($\Gamma_1 = 0.13, \Gamma_2 =0$), (right panel) Model III with  $\hat V_{\rm int} \propto \mathbb{I}$ ($\Gamma_1=0,\Gamma_2=0.15$).
  The blue dots denote the positions of the Dirac points of a pair of twisted but decoupled sheets. The actual Dirac points at $E=0$ are
  displaced from these points in Models II and III due to gauge potentials generated by the interlayer coupling. }
\end{center}
\end{figure}

\end{widetext}

\begin{widetext}

\begin{figure}
\begin{center}
  \includegraphics[angle=0,width=0.8\columnwidth]{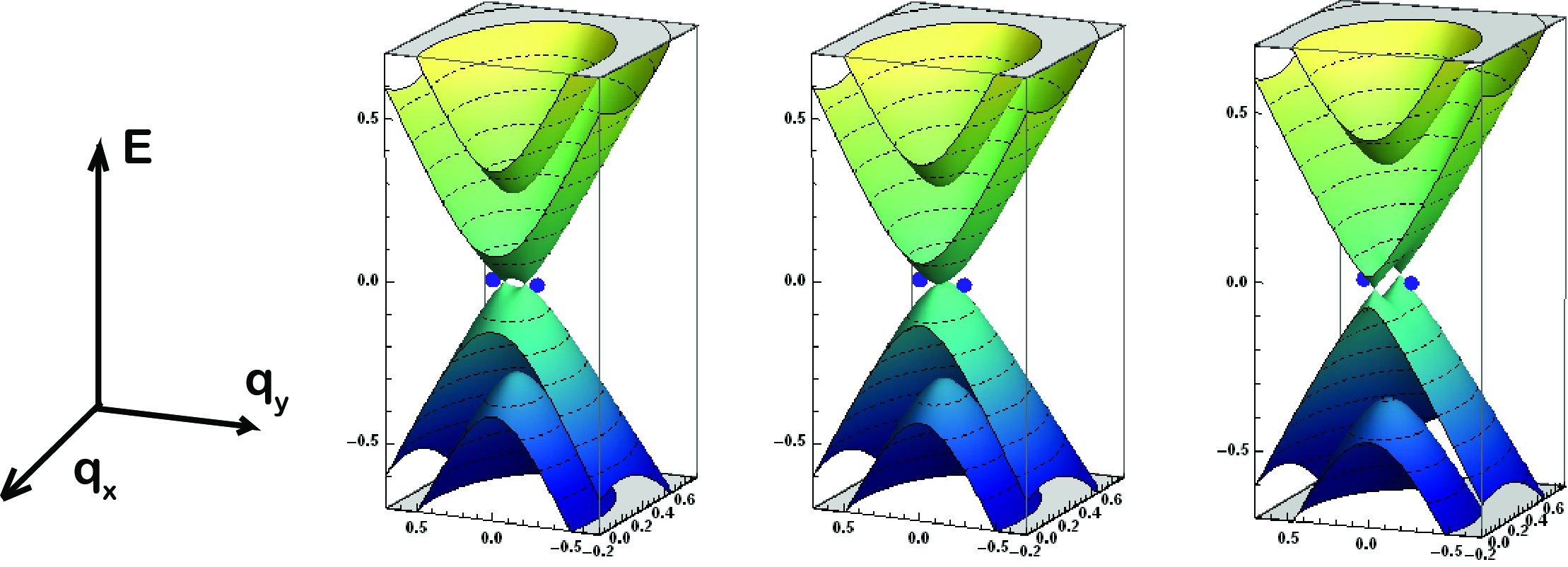}
  \caption{\label{bandsurfaceplotsII} Energy surfaces showing a transition from weak to strong coupling in interaction Model
  II with $\hbar v_F \Delta K = 0.15$. As the coupling parameter $\Gamma_1$ is increased the renormalized Dirac points evolve from weak coupling (left panel:
$\Gamma_1 = 0.13,\Gamma_2 =0$) and merge at a critical point (central
  panel: (left panel: $\Gamma_1 = 0.15,\Gamma_2 =0$)) and produce two new Dirac points in the strong coupling phase of the model (right panel: (left panel: $\Gamma_1 = 0.20,\Gamma_2 =0$)). The blue dots denote the Dirac
  points for a pair of twisted by decoupled layers. }
\end{center}
\end{figure}
\end{widetext}

In the opposite limit ($\Gamma_1=0,\Gamma_2 \neq 0$) one finds two low energy features that have the same helicity and repel
each other in momentum space so that no such $E=0$  merger occur regardless of the coupling strength $\Gamma_2$. Instead, at
{\it finite} energy the two Dirac cones cross to produce  new symmetry protected points which are second generation Dirac
singularities near $\pm \Delta$ as shown in the right hand panel of Figure \ref{bandsurfaceplotsI}. Despite the appearance of a
symmetry-allowed band crossing at the midpoints between the $E=0$ Dirac nodes, the system retains a saddle point structure
where the density of states is enhanced in an extended region of momentum space that is {\it laterally displaced} away from the
position of the crossing point. The low energy features of this type of coupling are quite distinct from those of Model I and
II. Notably, a perturbative velocity renormalization of the low energy Dirac nodes is symmetry forbidden for this class of
interaction models. This can provide an explananation for the absence of a velocity renormalization inferred from Landau level
spectroscopy \cite{STM} and from ARPES \cite{Hicks}, both of which find a Dirac velocity for twisted BLG that is the same as
the velocity for single layer graphene within experimental error. It also provides a interpretation for ARPES experiments that
clearly show a {\it crossing} of the Dirac cones cross instead of a hybridization induced anticrossing even in the limit of low
rotation angles \cite{Hicks}.

The spectra displayed in Figure \ref{bandsurfaceplotsI} and \ref{bandsurfaceplotsII}  show a distinct low energy topology
controlled by the symmetry of the long wavelength interlayer coupling matrix.  The singular points in these spectra have a
striking signature in the simulated ARPES momentum distributions as discussed below. Additionally there are special lines in
momentum space where ARPES selection rules are operative and these can be used to discriminate between these models and even to
directly identify the momentum shifts of their Dirac features due to the gauge coupling from the interlayer potential
\cite{Guinea,MelePRB}. The details of this analysis are presented in the following sections.

\section{ARPES Momentum Distributions}

In this section we discuss the constant energy momentum distributions calculated for the three limiting models discussed in
Section III, presented both as unweighted distributions and as distributions that are weighted by their photoemission
transition probabilities.

The top panel of Figure \ref{isocontours} shows the unweighted momentum distribution for the fully symmetric model with
($\Gamma_1 = \Gamma_2 =\Gamma$) where the interlayer matrix has the form $\hat V_{\rm int} =  \Gamma (\mathbb{I} + \sigma_1)$
(Model I). At energies well below the interlayer scale  $-\hbar v_F \Delta K$ the constant energy contours are self intesecting
loops. The inner loop maps out a constant energy contour on the lowest valence band and as the energy is increased (i.e. as it
is made less negative) this inner loop collapses to a point and vanishes at the extremum of the lowest band. The remaining
single surface develops a ``bean" shape that encircles both Dirac points highlighted by the bold dots. At still higher energy
this remaining contour fissions at energy of the saddle point producing two new closed curves that separately encircle the two
Dirac points. At positive energy these constant energy contours expand and reconnect at a positive saddle point energy to form
a single surface. At still higher energy these surface ``folds" contacting at an energy corresponding to the minimum of its
highest energy band leading again to a constant energy surface in the form of a self intersecting loop.  In the presence of an
interlayer bias (shown in the far right panel) the surfaces maintain the same topology but are inflated/deflated around the two
Dirac points in response to the layer symmetry breaking potential.

\begin{widetext}

\begin{figure}
\begin{center}
  \includegraphics[angle=0,width=0.8\columnwidth]{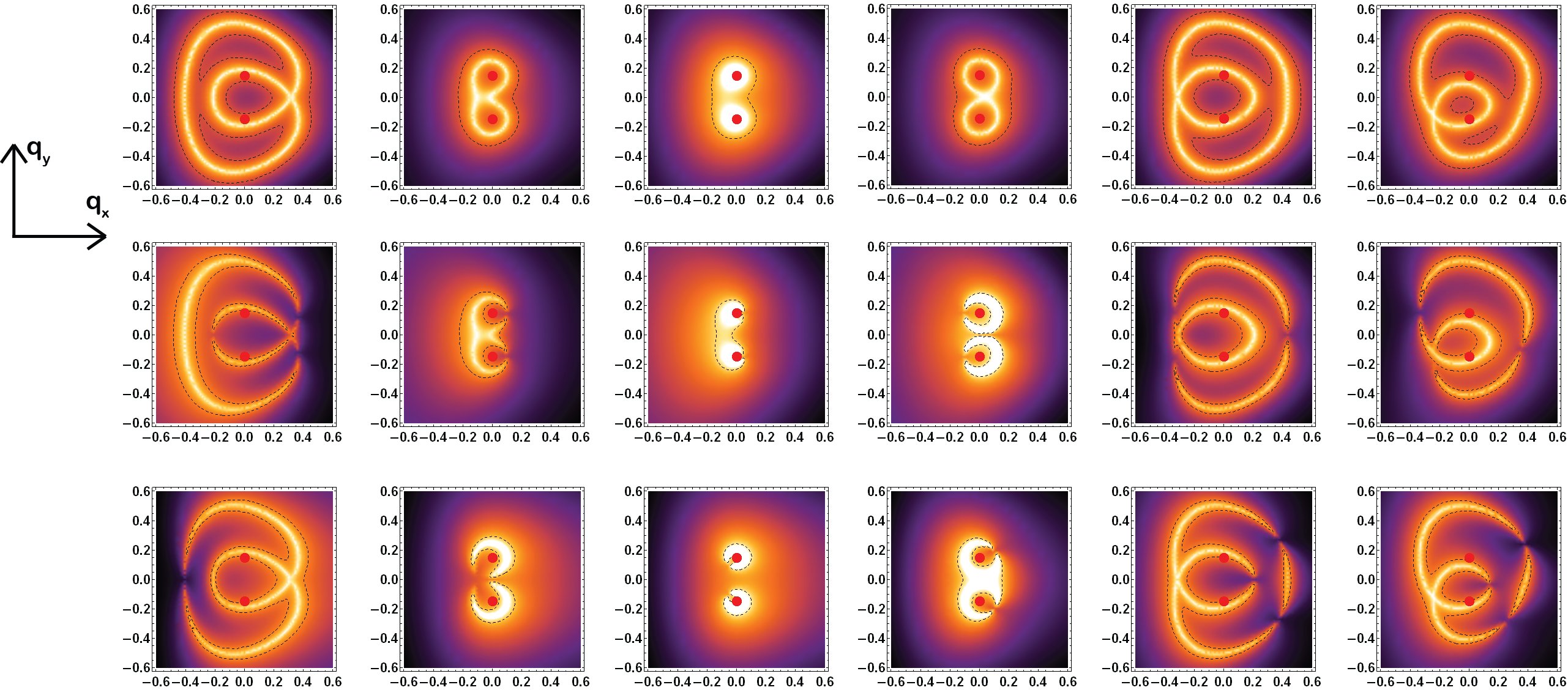}
  \caption{\label{isocontours} (Top panel) Evolution of the constant energy surfaces for the
  interaction Model I ($\hbar v_F \Delta K = 0.15, \Gamma_1=.05,\Gamma_2=.05$) taken at energies $-0.35, -0.12, -0.07, 0.12, 0.35$.
  The far right panel is the result with a interlayer bias.
  (Middle panel) Momentum distributions weighted by the transition probablities for $x$ polarized light.
  (Bottom panel) Momentum distributions weighed by the transition probabilities for $y$ polarized light. }
\end{center}
\end{figure}

\end{widetext}

The lower two panels of Figure \ref{isocontours} shows these distributions at the same sampling energies but weighted by the
squared modulus of their transition matrix elements for photoemission using light that is $x$ polarized (top row) and $y$
polarized (bottom row). For either polarization we observe a modulation of the simulated emission intensities the emergence of
a pattern of null points where the probability for photoemission vanishes. A symmetry analysis of these nodal surfaces is
presented in Section V. A comparison of these density plots shows that the intensity patterns undergo a complex evolution as a
function of energy with the nodal points undergoing pairwise creation and annihilation on constant energy surfaces in each band
at critical values of the initial state energy. As shown in the rightmost two panels, these null points are robust features of
the simulated ARPES momentum distributions and occur even when the layer symmetry is broken by an interlayer potential.

The top panel of Figure \ref{xcontours} shows the unweighted momentum distributions using the interlayer coupling model $\hat
V_{\rm int} = \Gamma_1 \sigma_x$ (Model II). This describes an interlayer Hamiltonian that is dominated by its hopping terms
that connect {\it different} sublattices, a situation that is  very nearly realized by the SWMc parameterization. This breaks
the symmetry of Model I, so that there are two separate (nonintersecting) constant energy surfaces at energies below $-\hbar
v_F \Delta K$. As the energy increases the innermost loop collapses to a point and vanishes (not shown) identifying the band
extremum of the lowest energy occupied band. At higher energy the remaining ring shrinks and fissions at a saddle point energy
forming two loops which in turn collapse around the two layer-coupled Dirac points singularities. As noted in Section III these
points are generally displaced from the unperturbed Dirac points of the two decoupled layers due to a gauge potential generated
by the interlayer coupling. In the weak coupling limit (i.e. for large rotation angle) these new point singularities are
located along the vertical axis of these plots. As the interaction strength is increased the system these singular points merge
and the system undergoes a topological transition to a new strong coupling regime where the singular points are symmetrically
positioned along the horizontal axis.

\begin{widetext}

\begin{figure}
\begin{center}
  \includegraphics[angle=0,width=0.8\columnwidth]{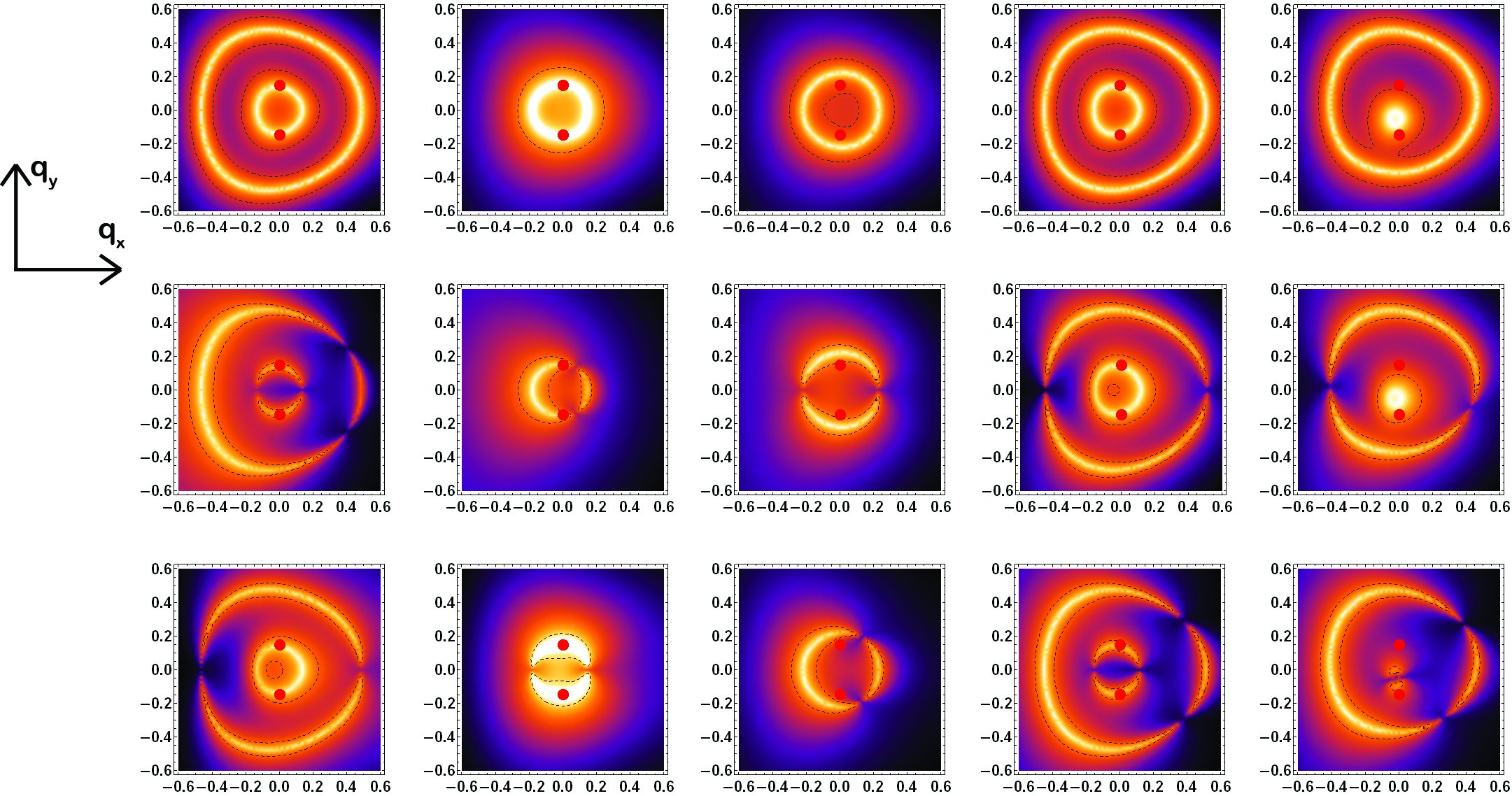}
  \caption{\label{xcontours} (Top panel) Evolution of the constant energy surfaces for the
  interaction Model $V_{\rm int} \propto \sigma_x$ ($ \hbar v_F \Delta K = 0.15, \Gamma_1=0.13,\Gamma_2=0.0$) at energies ($-0.35, -0.07, 0.12, 0.35$)
  with energy values (left to
  right). These parameters describe this model in the weak coupling limit, i.e. below its topological transition.
  The far right panel is the result with an interlayer bias.
  (Middle panel) Momentum distributions weighted by the transition probablities for $x$ polarized light.
  (Bottom panel) Momentum distributions weighed by the transition probabilities for $y$ polarized light. }
\end{center}
\end{figure}

\end{widetext}

The lower two panels of Figure \ref{xcontours} show the probablity weighted distributions calculated for Model II for the same
representative energies. Again we observe a complex modulation of the emission intensities with null points identified in the
constant energy surfaces for both polarizations. The locations of the nodal points in these surfaces exhibit an interesting
oscillation as a function of band index and light polarization. Note for example the interchange of the positions of the
emission maxima and nodal points in the second and third bands under exchange of the $x$ and $y$ polarization.

\begin{widetext}

\begin{figure}
\begin{center}
  \includegraphics[angle=0,width=0.8\columnwidth]{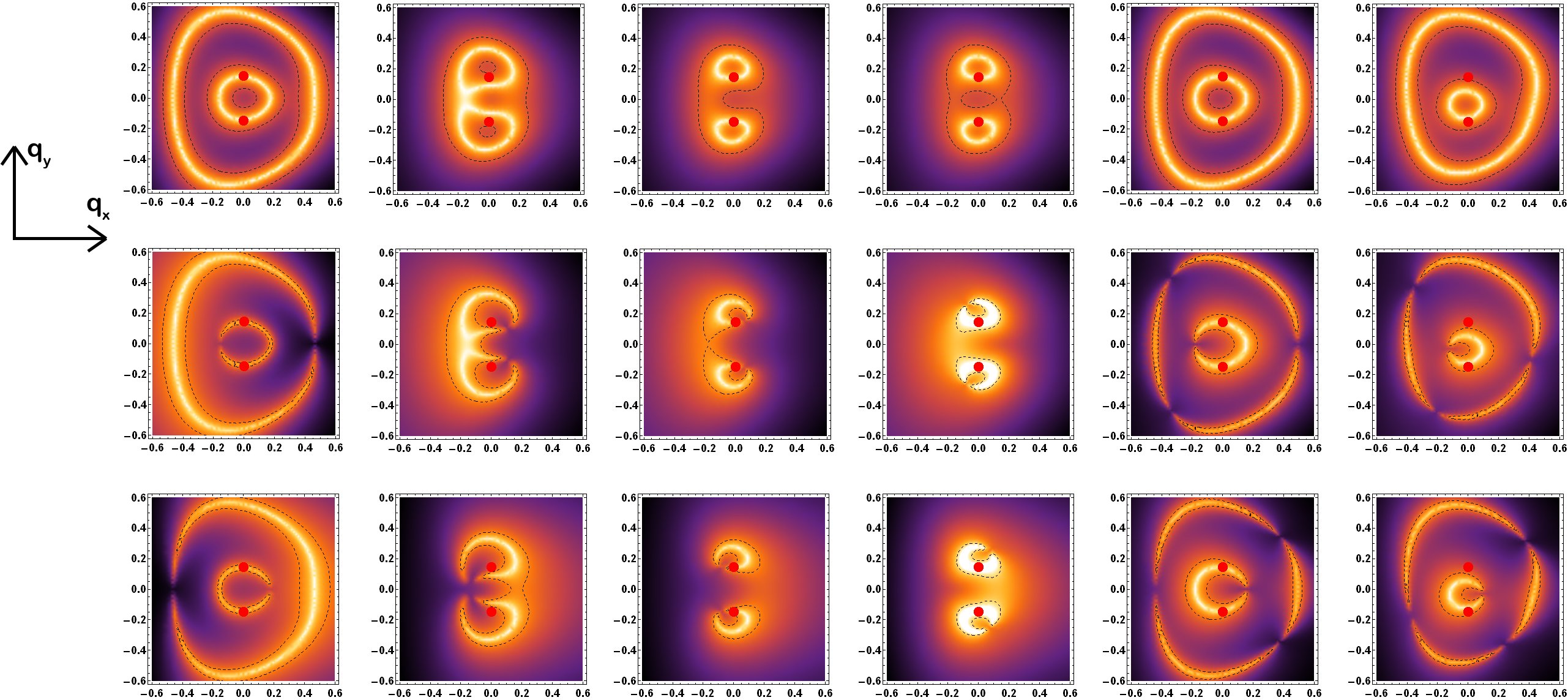}
  \caption{\label{Icontours} (Top panel) Evolution of the constant energy surfaces for the
  interaction Model $V_{\rm int} \propto {\mathbb I}$ ($\hbar v_F \Delta K = 0.15, \Gamma_1=0,\Gamma_2 = 0.15$) at energy values ($-0.35, -0.12, -0.07, 0.12, 0.35$).
  The far right panel is the result with an interlayer bias.
  (Middle panel) Momentum distributions weighted by the transition probablities for $x$ polarized light.
  (Bottom panel) Momentum distributions weighed by the transition probabilities for $y$ polarized light.}
\end{center}
\end{figure}

\end{widetext}

Figure \ref{Icontours} presents the unweighted momentum distributions obtained from a similar analysis using a coupling model
of the form$\hat V_{\rm int} = \Gamma_2 \sigma_0$ (Model III). In this limit the interlayer coupling is dominated by amplitudes
connecting sites on the {\it same} sublattice. A coupling of this type would occur in a putative $AA$ stacked geometry which,
though apparently excluded by the conventional SWMc parameters of Table I, would provide an explanation for the absence of a
Fermi velocity renormalization and the intersection of Dirac bands observed in twisted graphenes grown on SiC $(000 \bar 1)$.
At low energy the constant energy surfaces form two disconnected loops. The signature of coupling in the form of Model III is
the collapse of the inner ring at critical energy identifying second generation Dirac point of Figure \ref{bandsurfaceplotsI}
and its re-emergence as a closed loop immediately above this energy (not shown). As the energy is increased still further this
ring bifurcates at a pair of saddle points positioned displaced along the horizontal ($q_x$) axes. These two surfaces collapse
around the two $E=0$ Dirac nodes in the model. Here, because of the gauge potential generated by the interlayer coupling the
$E=0$ nodes are repelled from the geometrically defined Dirac points. As one crosses to positive energies this evolution is
reversed leading to the reconstruction of two ringlike surfaces for large positive energies.

The lower two panels of Figure \ref{Icontours} show the distributions of Model III weighted by their transition probabilities.
Again we observe the complementary positions of the emission maxima and nodal points from a single band for $x$ and $y$
polarized excitation.

\section{Symmetry Analysis}

The nodal patterns identified in Section IV can be understood by a consideration of the analytic structure of the complex
transition amplitudes ${\cal M}({\bm p}, {\bm q})$ in Eqn. (\ref{M}). The key observation is that in particular circumstances
identified below this complex transition amplitude degenerates to a real function of ${\bm q}$. In this situation ${\cal M}$
will generically vanish along {\it lines} in reciprocal space (where it changes sign) rather than at points (where it wraps its
phase). The evolution of the nodal patterns in ARPES distributions then correspond to the evolution of constant energy surfaces
through a network of lines along which ${\cal M}$ is zero. These lines can terminate at singular points in the band structure.
In the simplest picture, these points can be identified with the rotationally offset Dirac points. The actual situation is
richer however since, as we have seen, these points are generally displaced from the rotationally defined Dirac points by gauge
potentials produced by the interlayer coupling and in addition second generation symmetry protected singularities can appear
elsewhere the band structure. The identification of these lines, their continuity and termination points in ${\bm q}$ space
provides a powerful probe of the nature of the interlayer coherence.

\subsection{Intralayer Interference of the Transition Amplitudes}\label{symmetryintra}

The layer diagonal blocks in Eqn. \ref{HamT} are momentum-displaced and -rotated versions of the Hamiltonian \ref{DiracS} and
have the symmetry
\begin{eqnarray}\label{symmetry}
{\cal H}_S = \sigma_x {\cal H}_S^* \sigma_x
\end{eqnarray}
so that the eigenfunctions are equal weight states
\begin{eqnarray}\label{chi}
\chi({\bm q}) = \frac{1}{\sqrt{2}} \left(%
\begin{array}{c}
  e^{-i \alpha ({\bm q})} \\
  \pm e^{i \alpha ({\bm q})} \\
\end{array}%
\right)
\end{eqnarray}
The velocity operators of Eqn. \ref{velocities} are layer block-diagonal and $\theta$-dependent and can be represented
compactly
\begin{eqnarray}\label{rotv}
v_\mu = \left(%
\begin{array}{cc}
  1 & 0 \\
  0 & e^{i \theta/2} \\
\end{array}%
\right) \sigma_\mu \left(%
\begin{array}{cc}
  1 & 0 \\
  0 & e^{-i \theta/2} \\
\end{array}%
\right)
\end{eqnarray}
in the top layer with positive rotation angle $\theta$ where $\sigma_\mu$ is the $\mu$-th $2 \times 2$ Pauli matrix. Reversing
the sign of $\theta$ in this expression gives the related velocity operator projected on the bottom layer. Combining Eqns.
(\ref{chi}) and (\ref{rotv}) we identify the transition states
\begin{eqnarray}
\chi^\pm_x ({\bm q}) = v_x \cdot \chi ({\bm q}) = \frac{1}{\sqrt{2}} \left(%
\begin{array}{c}
  \pm e^{i (\alpha ({\bm q})-\theta/2)} \\
   e^{ - i (\alpha ({\bm q})-\theta/2)} \\
\end{array}%
\right) \,\, {\rm x-polarized} \nonumber\\
\chi^\pm_y ({\bm q}) = v_y \cdot \chi ({\bm q}) = \frac{i}{\sqrt{2}} \left(%
\begin{array}{c}
  \mp e^{i (\alpha ({\bm q})-\theta/2)} \\
   e^{-i (\alpha ({\bm q})-\theta/2)}  \\
\end{array}%
\right) \,\, {\rm y-polarized} \nonumber\\
\end{eqnarray}
The photoemission matrix elements obtained by projecting these transition states onto the plane wave final state \ref{pw} for
$x$ polarized excitation are
\begin{eqnarray}\label{xpol}
{\cal M}^+_x({\bm p}, {\bm q}) = \frac{1}{\sqrt{2}}
\cos (\alpha ({\bm q}) - \theta/2) \nonumber\\
{\cal M}^-_x({\bm p}, {\bm q})  = \frac{-i}{\sqrt{2}}
\sin (\alpha ({\bm q}) - \theta/2) \nonumber\\
\end{eqnarray}
 and for $y$-polarized excitation
\begin{eqnarray}\label{ypol}
 {\cal M}^+_y({\bm p}, {\bm q})  =
\frac{1}{\sqrt{2}}
\sin (\alpha ({\bm q}) - \theta/2) \nonumber\\
{\cal M}^-_y({\bm p}, {\bm q})  = \frac{i}{\sqrt{2}}
 \cos (\alpha ({\bm q}) - \theta/2) \nonumber\\
\end{eqnarray}

The structure of these matrix elements has been studied previously in the context of single layer graphene. Note the reversal
of the ${\bm q}$-space modulations the of $x$ and $y$ polarized transition amplitudes for the positive and negative energy
eigenstates. Importantly the transition amplitudes vanish along {\it lines} where $\alpha({\bm q}) = \theta/2$ or $\alpha({\bm
q}) = \theta/2 \pm \pi/2$. These nodal lines terminate at points where the layer projected Hamiltonian  $\propto {\mathbb I}$
and $\chi^+$ and $\chi^-$ are degenerate.  Crossing this contact point reverses the assignment of $\pm$ indices to the upper
and lower energy bands, thereby terminating the nodal line in a single band and transferring a nodal line to its partner. This
phenomenon is responsible for the one sided ``dark corridor" seen in photoemission from single layer graphene
\cite{Kern,Louie}. Equations \ref{xpol} and \ref{ypol} are rotated versions of that result where the new dark corridors are
rotated at angles $\pm \theta/2$ in the two layers due to the rotational misalignment.

Any real combination of the transition amplitudes in Eqns. (\ref{xpol}) and (\ref{ypol}) describes linearly polarized
excitation at some general angle of polarization. We see that its effect is simply to rotate the nodal line for photoemission
while preserving an endpoint anchored at a singular point of degeneracy. By contrast elliptically or circularly polarized light
requires a phase lag between two orthogonal polarizations of the radiation. In this situation the transition matrix element is
necessarily represented by a two dimensional parameter space (it must be a complex scalar function) and it can vanish only at
special points in momentum space where two independent transition amplitudes vanish.  Such points would be difficult to measure
in ARPES since this would require fine tuning the initial state to a single critical energy.

These conclusions hold quite generally for any system described by the symmetry of Eqn.  (\ref{symmetry}) in its layer
projected Hamiltonians. For coupled BLG, if the interlayer coupling matrix $\hat V$ respects sublattice symmetry (as occurs in
models in the form of Eqn. {\ref{HamT}),  one can obtain an effective layer projected theory that respects sublattice symmetry
by integrating out one of the layers. Such a system generically exhibits one-sided line nodes in the photoemission matrix
elements. Since the eigenfunctions of this layer projected Hamiltonian are equal weight states of the form given in Eqn.
(\ref{chi}) its matrix elements are controlled by the evolution of a single scalar parameter $\alpha({\bm q})$. This requires
that its nodal lines are continuous except at point singularities where $\nabla \alpha$ is undefined and a nodal line can
either terminate or switch between bands that are degenerate at the singularity.  As noted in Section IV the endpoints of these
nodal lines will generally deviate from the rotationally defined points of degeneracy if a residual gauge potential  is induced
by integrating out the interlayer coupling. Nonetheless the physics responsible for this behavior is essentially the singular
structure of the single layer Hamiltonian perturbatively renormalized by interactions between layers.

\subsection{Interlayer Interference of the Transition Amplitudes}\label{symmetryinter}

Transition matrix elements derived from the Hamiltonian in Eqn. (\ref{HamT}) support additional interference features that
arise directly from its interlayer coupling amplitudes. These provide a more sensitive probe of the symmetry of the interlayer
Hamiltonian in twisted BLG and are analyzed in this section.

Our approach exploits a symmetry in Eqn. (\ref{HamT}) along the line $q_y=0$  where because of the reversal of the momentum
offsets $\pm \Delta {\bm K}/2$ the layer-projected Hamiltonians in the top and bottom layers are related
\begin{eqnarray}\label{topbottom}
{\cal H}_{\rm bottom} = {\cal H}^*_{\rm top} \nonumber\\
\end{eqnarray}
This symmetry is broken for a general two dimensional momentum ${\bm q}$ but it remains a local symmetry for all momenta $q_x$
along the midline. Along this line the eigenfunctions are equal-weight combinations of the four layer-orbital degrees of
freedom. Additionally, since the Hamiltonian commutes with the operator $R = \tau_x \sigma_x$, these eigenstates can be indexed
by their parities ($\pm$) under $R$ and take the form
\begin{eqnarray}
\Psi^\pm ({\bm q}) = \frac{1}{2} \left(%
\begin{array}{c}
  e^{-i \alpha({\bm q})} \\
   e^{i \alpha({\bm q})} \\
   \pm e^{i \alpha ({\bm q})} \\
   \pm e^{-i \alpha({\bm q})} \\
\end{array}%
\right)
\end{eqnarray}

The four-component velocity operators are
\begin{widetext}
\begin{eqnarray}\label{rotv2}
V_\mu = \left(%
\begin{array}{cccc}
  1 &   &   &  \\
   & e^{i \theta/2} &  &  \\
    &   & 1 &  \\
   &  &  & e^{-i \theta/2} \\
\end{array}%
\right)
\left(%
\begin{array}{cc}
  \sigma_\mu & 0 \\
  0 & \sigma_\mu \\
\end{array}%
\right) \left(%
\begin{array}{cccc}
  1 &   &   &  \\
   & e^{-i \theta/2} &  &  \\
    &   & 1 &  \\
   &  &  & e^{i \theta/2} \\
\end{array}%
\right)
\end{eqnarray}
\end{widetext}
and the transition states are obtained by application of these operators to the initial states $\Psi^\pm$, giving for $x$
polarized light:
\begin{eqnarray}
\Psi^\pm_x = V_x \cdot \Psi^\pm = \frac{1}{2} \left(%
\begin{array}{c}
  e^{i(\alpha-\theta/2)} \\
  e^{-i(\alpha-\theta/2)} \\
  \pm e^{-i(\alpha-\theta/2)} \\
  \pm e^{i(\alpha-\theta/2)} \\
\end{array}%
\right)
\end{eqnarray}
and for $y$ polarized light
\begin{eqnarray}
\Psi^\pm_y = V_y \cdot \Psi^\pm = \frac{1}{2} \left(%
\begin{array}{c}
  e^{i(\alpha-\theta/2)} \\
  -e^{-i(\alpha-\theta/2)} \\
  \pm e^{-i(\alpha-\theta/2)} \\
  \mp e^{i(\alpha-\theta/2)} \\
\end{array}%
\right)
\end{eqnarray}
The matrix elements for photoemission in this case depends on the {\it relative phase} $\phi$ for the layer amplitudes in the
outgoing state in Eqn. (\ref{pw}). For $x$ polarized light the matrix elements are
\begin{eqnarray}\label{midlinemx}
{\cal M}^+_x ({\bm p}, {\bm q}) = \cos(\alpha-\theta/2) \cos(\phi/2) \nonumber\\
{\cal M}^-_x ({\bm p}, {\bm q}) = -i\cos(\alpha-\theta/2) \sin(\phi/2)
\end{eqnarray}
and for $y$ polarized light
\begin{eqnarray}\label{midlinemy}
{\cal M}^+_y ({\bm p}, {\bm q}) = \sin(\alpha-\theta/2) \sin(\phi/2) \nonumber\\
{\cal M}^-_y ({\bm p}, {\bm q}) = i \sin(\alpha-\theta/2) \cos(\phi/2)
\end{eqnarray}
In Eqns. (\ref{midlinemx}) and (\ref{midlinemy}) the phases $\alpha$ and $\theta$ are independent variables (the former is
determined by coefficients in the Hamiltonian and the latter is the rotation angle) so that generically $\alpha - \theta/2$ is
not a multiple of $\pi/2$. Thus the relevant interference physics is fully controlled by the interlayer phase $\phi$. When the
outgoing waves from the two layers are in phase ($\phi = 2 m \pi$) emission from states with even $R$-parity is allowed for $x$
polarized and forbidden for $y$ polarized light. Conversely, emission from the odd $R$-parity states is allowed for $y$
polarized light and forbidden for $x$ polarization. When the phase lag between layers $\phi = (2m+1) \pi$ these selection rules
are exactly reversed. As is the case for intralayer interference, nodal lines are absent for circular or elliptical
polarization since this requires a simultaneous vanishing of the matrix elements in two orthogonal excitation channels.

When $\phi \neq (0,\pi)$ the emission between layers is neither exactly in or out of phase and nodal lines do not occur along
the midline.  Instead they are replaced by troughs (local minima) in the ARPES momentum distributions. The $\phi$-derivatives
of the momentum distributions measured for two orthogonal linear polarizations allow one to extract the internal phases
$\alpha$ that define the wavefunctions $\Psi_\pm$ along this line. Physically these $\phi$ derivatives can be measured by
measuring the differential change of the emission intensity at each position in ${\bm q}$ space distribution as a function of
the excitation energy. Using Eqns. (\ref{midlinemx}) and (\ref{midlinemy}) one finds that the intensities for even $R$-parity
state in two orthogonal polarizations are

\begin{eqnarray}
I_x  &=& \cos^2(\alpha-\theta/2) \left( \frac{ 1 + \cos(\phi)}{2} \right) \nonumber\\
I_y &=& \sin^2(\alpha-\theta/2) \left( \frac{ 1 - \cos(\phi)}{2} \right)
\end{eqnarray}
and therefore the {\it ratio} $r_{yx} = - (dI_y/d \phi)/(dI_x/d \phi)$ is $\phi$ (energy) independent and allows one to
identify $\alpha ({\bm q})$
\begin{eqnarray}\label{alpha}
\alpha ({\bm q}) = \frac{\theta}{2} + \arctan \sqrt{r_{yx}({\bm q})}
\end{eqnarray}
Eqn. (\ref{alpha}) holds everywhere along the midline when the interlayer bias $V=0$. When the interlayer bias is nonzero a
similar expression can be used to determine the wavefunctions along a hyperbolic locus in ${\bm q}$-space where the
wavefunction $\Psi_\pm$ is an equal weight state shared between the two layers.  Illustrative examples are given in Section
V.D.

Eqns. (\ref{midlinemx}) and (\ref{midlinemy}) predict that nodal lines at $q_y=0$ extend over the entire midline when the $R$-
parity of the initial state is constant along this line. This is always the case when the bands are nonintersecting and
illustrations of this are given in the following sections. Note however that when a symmetry protected band {\it crossing}
occurs along the midline, the $R$-parity changes its sign at a singular point of degeneracy identifiable by the  termination a
nodal line in one band (ordered by its energy) and its appearance in another (also ordered by energy). The $q_y=0$ midline is
perpendicular to the offset $\Delta {\bm K}$ between the bare Dirac nodes of the two layers so that crossings of this type do
not occur in the simplest model of the rotated bilayer (Model I). However, in refined models for twisted bilayer graphene
symmetry protected crossings {\it do} occur, notably along the line  $q_y=0$ near $E=0$ in the strong coupling limit of Model
II and at the energy of the second generation Dirac point (near the interlayer scale $\hbar v_F \Delta K$) for any coupling
strength in Model III. This provides a unique spectroscopic diagnostic of singularities in their band structures.

The interlayer selection rules can be understood from the transformation of the current operators $V_\mu$ under the symmetry
operation $R$. $V_{x(y)}$ have even(odd) $R$-parities, and for $\phi = 0(\pi)$ the outgoing plane wave state in Eqn. (\ref{pw})
is even(odd). Thus the even parity initial state $\Psi^+$ can be excited into the outgoing final state with $\phi=0(\pi)$ only
using $x(y)$ polarization. Similar considerations apply for general interlayer phase difference $\phi$ with a rotation of the
principal axes as noted above.

\subsection{Branching Nodal Surfaces}

The nodal patterns discussed in Section \ref{symmetryintra} arise from cancellation of photoemission amplitudes from the
degrees of freedom in the individual layers, and in Section \ref{symmetryinter} they arise from cancellation of amplitudes from
different layers. There are special points in ${\bm q}$ space where both destructive interference conditions are satisfied.
These can be identified as branching points where a nodal line running along the midline $q_y=0$ develops branches that
continue to the (renormalized) Dirac points of the two layers. The locations of these branching points can be determined from
consideration of the bilayer coherent wavefunctions along the high symmetry $q_y=0$ midline and are analyzed in this section.

Using the Hamiltonian in Eqn. (\ref{HamT}) we partition the Hamiltonian at $q_y \neq 0$ into a Hamiltonian along the midline at
$q_y=0$ and a $q_y$ dependent piece
\begin{eqnarray}
{\cal H}(q_y) &=& {\cal H}(q_y=0) + {\cal H}'(q_y) \nonumber\\
&=& {\cal H}(q_y=0) + q_y \tau_0 \sigma_2
\end{eqnarray}
Therefore the eigenstates near the midline can be expanded in terms of the eigenstates along the midline
\begin{eqnarray}
\Psi_m(q_y) &=& \Psi_m (q_y=0)  \nonumber\\
 &+& \,\,\,\,  q_y \sum_n \, \Psi_n(0) \frac{\langle \Psi_n(0) | \sigma_2 \tau_0 | \Psi_m(0)
\rangle}{E_m - E_n}
\end{eqnarray}
The midline eigenfunctions are also eigenfunctions of $R= \sigma_x \tau_x$, and since ${\cal H}'$ is odd under $R$  the sum
connects only states of opposite $R$-parity. In a similar way the photoemission matrix element for the $m$-th band in the
$\mu$-th polarization ${\cal M}_\mu$ can be expanded
\begin{eqnarray}\label{branchmid}
{\cal M}_{m,\mu} (q_y) &=& {\cal M}_{m,\mu} (0)  \nonumber\\
 &+& \,\,\,\, q_y  \sum_n \frac{\langle \psi_{\bm p}^> | j_\mu |\Psi_n(0) \rangle  \langle \Psi_n(0) | \sigma_2 \tau_0 | \Psi_m(0)
\rangle}{E_m - E_n} \nonumber\\
\end{eqnarray}
The sum in Eqn. (\ref{branchmid}) runs over the two states that reverse the $R$ parity of the initial state $\Psi_m(0)$ and
using Eqns. (\ref{midlinemx}) and (\ref{midlinemy}) when ${\cal M}_{m,\mu} (0)=0$ one sees that ${\cal M}_{n,\mu} (0)$ is
nonzero for each of these admixed states. This means that the photoemission matrix element turns on linearly as a function of
the transverse momentum $q_y$ and the nodal surface is an unbranched line for a general momentum $q_x$.

However a singular point  can occur along the midline  $(q_b,0)$ where the two contributions in the sum cancel. To locate such
a point we label the two energy denominators between the state $m$ and its two intermediate states $n_1$ and $n_2$ by $\Delta_1
= E_m - E_{n_1}$ and $\Delta_2 = E_m - E_{n,2}$. Then the $q_y$-linear changes to the matrix element can be written as a real
function of the phase angle $\alpha$ giving
\begin{eqnarray}\label{diffM}
\delta {\cal M} &=& q_y \left( \frac{\cos 2\alpha \, \cos (\alpha - \theta/2)}{\Delta_1} + \frac{\sin 2\alpha \, \sin (\alpha -
\theta/2)}{\Delta_2} \right) \nonumber\\
 &=& B q_y \cos(\beta - \alpha + \theta/2)
\end{eqnarray}
where $\beta = \arctan( (\Delta_1/\Delta_2) \tan 2 \alpha)$ and $B$ is a constant. A critical branching point occurs when the
right hand side of Eqn. (\ref{diffM}) is zero. A nodal line forms two branches above and below the $q_x$ axis at such a
critical point, once formed these arms continue to the Dirac points along the $q_y$ axis. This can be understood as a
consequence of the continuity of the intralayer nodal lines that must connect the two Dirac points on opposite sides of the
$q_x$ axis.

\subsection{Examples}

\subsubsection{Model I}

The nodal lines for Model I with $\hat V_{\rm int} \propto (\mathbb{I} + \sigma_x)$ are shown in Figure \ref{Model3Nodes}. We
observe that the nodal lines that occur along the midline $q_y=0$ propagate without termination. This occurs because the
presence or absence of a nodal line is determined by the polarization state of the light and by the $R$-parity of the state,
which does not change as a function of $q_x$ along the midline. The interlayer nodal lines also occur as complementary pairs,
i.e. they are present for a particular band in $x$ polarization only when absent in $y$ polarization and vice versa. This
occurs whenever the phase lag $\phi$ for emission between layers is a multiple of $\pi$. The phase difference $\phi$ can be
tuned continuously by varying the excitation energy and for $\phi \neq m \pi$ the nodal lines are absent in both $x$ and $y$
polarizations. However for intermediate $\phi$ these lines are recovered when the principal polarization axes are also rotated
according to the formulas given in Eqns. (\ref{midlinemx}) and (\ref{midlinemy}). No such energy dependence occurs for the
nodal surfaces that terminate on the Dirac points which arise essentially from intralayer interference in the transition
amplitudes and are therefore $\phi$-independent. The energy and polarization dependence of the nodal lines can therefore be
used as a smoking gun to discriminate between intralayer and interlayer interference phenomena in the momentum distributions.

\begin{widetext}

\begin{figure}
\begin{center}
  \includegraphics[angle=0,width=0.8\columnwidth]{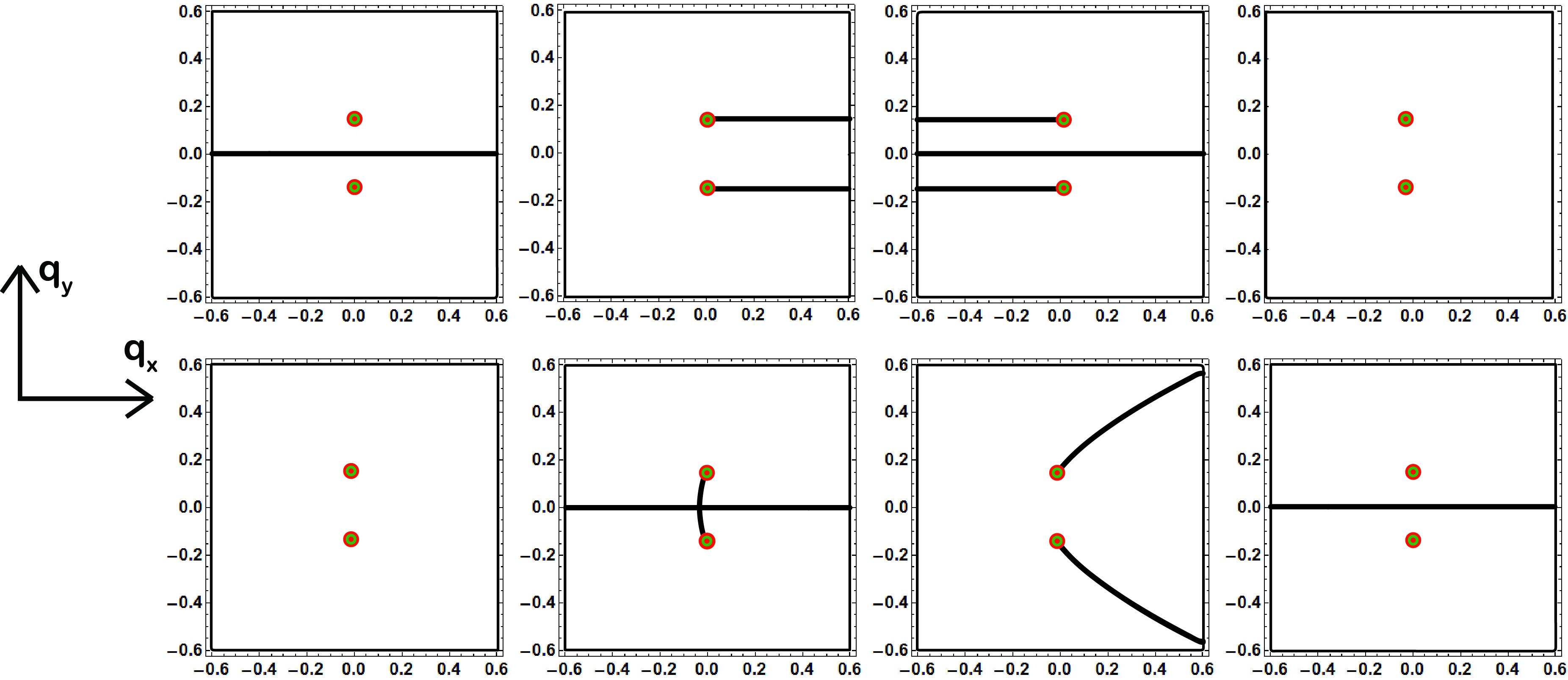}
  \caption{\label{Model3Nodes} Nodal lines in the photoemission amplitudes calculated for interlayer coupling Model I with $\hat V_{\rm int} \propto (\mathbb{I} + \sigma_x)$.
  The top row is for $x$-polarized light and the bottom row is for $y$ polarized light. The four columns are for the four bands
  of the model ordered by energy. The $x$ polarized amplitudes in bands 2 and 3 have one-sided nodal lines at constant values of $q_y$ while the $y$ polarized amplitudes
  have nodal curves due to a gauge potential generated by the interlayer coupling. In bands 2 and 3 the nodal lines terminate at points (red dots)
  that can be identified with the Dirac points of the decoupled layers (green dots).}
\end{center}
\end{figure}

\end{widetext}

There are additional nodal lines in the second and third bands that terminate at the Dirac points. The termination and the
appearance of these features correspond to a transferring a line node between the bands that touch at the Dirac points.The
signature of the nodal patterns produced by the coupling in Model I is the appearance, in $x$ polarized excitation, of
one-sided horizontal nodal lines that terminate at the Dirac points. This occurs because the interlayer operator in Model I has
one zero eigenvalue, implying that there is one state with a particular pseudospin polarization in each layer that cannot be
transported to its neighboring layer. This state is the antisymmetric eigenfunction of $\sigma_x$ and it cannot be coupled to
the outgoing $\sigma_x$-symmetric final state by $x$ polarized radiation. Since these states are localized to the individual
layers, they are eigenfunctions of the single layer $2 \times 2$ Dirac operators for which the pseudospin polarization remains
constant along radially directed lines in $\bm{q}$ space. Thus there are two such states with nodal lines that terminates at
the {\it unperturbed} Dirac nodes of the two layers. In $y$ polarization the related lines nodes share the same termination
points but the pseudospin polarization is a function of both $q_x$ and $q_y$ and is constant along a curve in momentum space,
as shown in the lower panel. The curvature of this line is a manifestation of the momentum dependence of the gauge potential
produced by the interaction with the neighboring layer. The termination points for these lines are the same for $x$ and $y$
polarized excitation, and thus the point singularities in this model can be identified with the geometrically determined Dirac
points of two decoupled layers. This is a special feature of this class of interaction models.

\subsubsection{Model II}

In Model II $\hat V_{\rm int} \propto \sigma_x$ so that the interlayer coupling is controlled by terms that connect opposite
sublattice sites on neighboring layers. The signature of this model is the existence of a weak and strong coupling limit
separated by a critical state where the renormalized Dirac points of the two layers are merged. The nodal structure in these
two regimes is illutrated in Figures \ref{Model2subNodes} and \ref{Model2supNodes}. The dimensionless coupling strength $\tilde
\Gamma_1 = c_{AB}/\hbar v_F \Delta K$ so that the strong coupling limit can be realized for sufficiently small rotation angles.
Using the data of Table I, one expects the strong coupling regime to describe the physics for rotation angles $\theta < 4^{\rm
\circ}$.

\begin{widetext}

\begin{figure}
\begin{center}
  \includegraphics[angle=0,width=0.8\columnwidth]{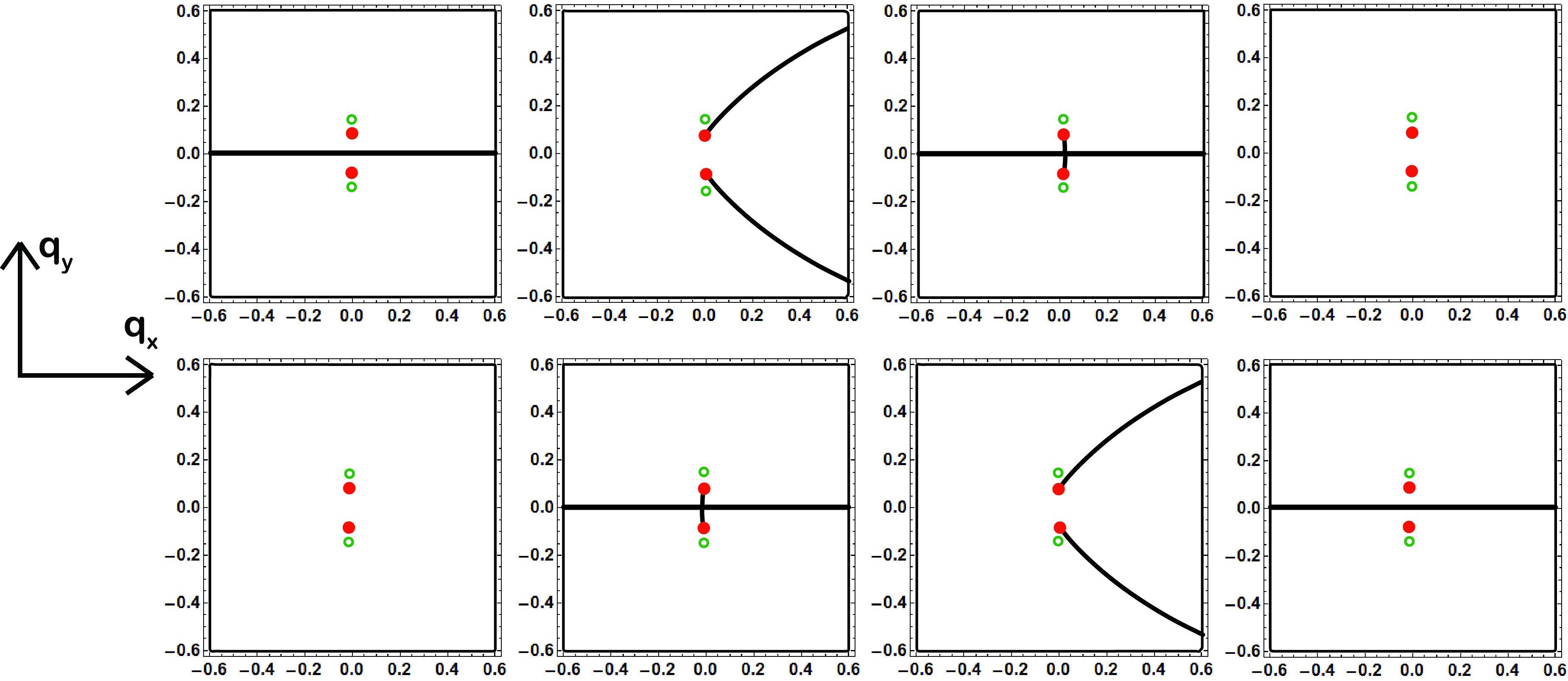}
  \caption{\label{Model2subNodes} Nodal lines in the photoemission amplitudes calculated for interlayer coupling Model II with $\hat V_{\rm int} \propto  \sigma_x$.
  The top row is for $x$-polarized light and the bottom row is for $y$ polarized light. The four columns are for the four bands
  of the model ordered by energy. The data are shown for Model II in its weak coupling regime before the merger and reconstruction of
  the $E=0$ Dirac points. The renormalized Dirac points (red dots) are displaced from the geometrically defined Dirac points of
  two decoupled layers (open green circles). The nodal lines in bands 2 and 3 terminate on the renormalized Dirac points. An
  additional nodal line along the midline is produced by  interlayer interference of the transition amplitudes. }
\end{center}
\end{figure}

\end{widetext}

The nodal structure for Model II in the weak coupling regime is shown in Figure \ref{Model2subNodes}. It shares some features
in common with Model I and the nodal pattern in $y$ polarization is quite similar. The main difference is in $x$ polarization
where ``flat" one-sided nodal lines of Model I evolve into dispersive nodal curves in Model II. As noted above the curvature of
these lines is a momentum-space manifestation of the interlayer gauge potential for the twisted bilayer which is absent by
symmetry for a single polarization in Model I.

\begin{widetext}

\begin{figure}
\begin{center}
  \includegraphics[angle=0,width=0.8\columnwidth]{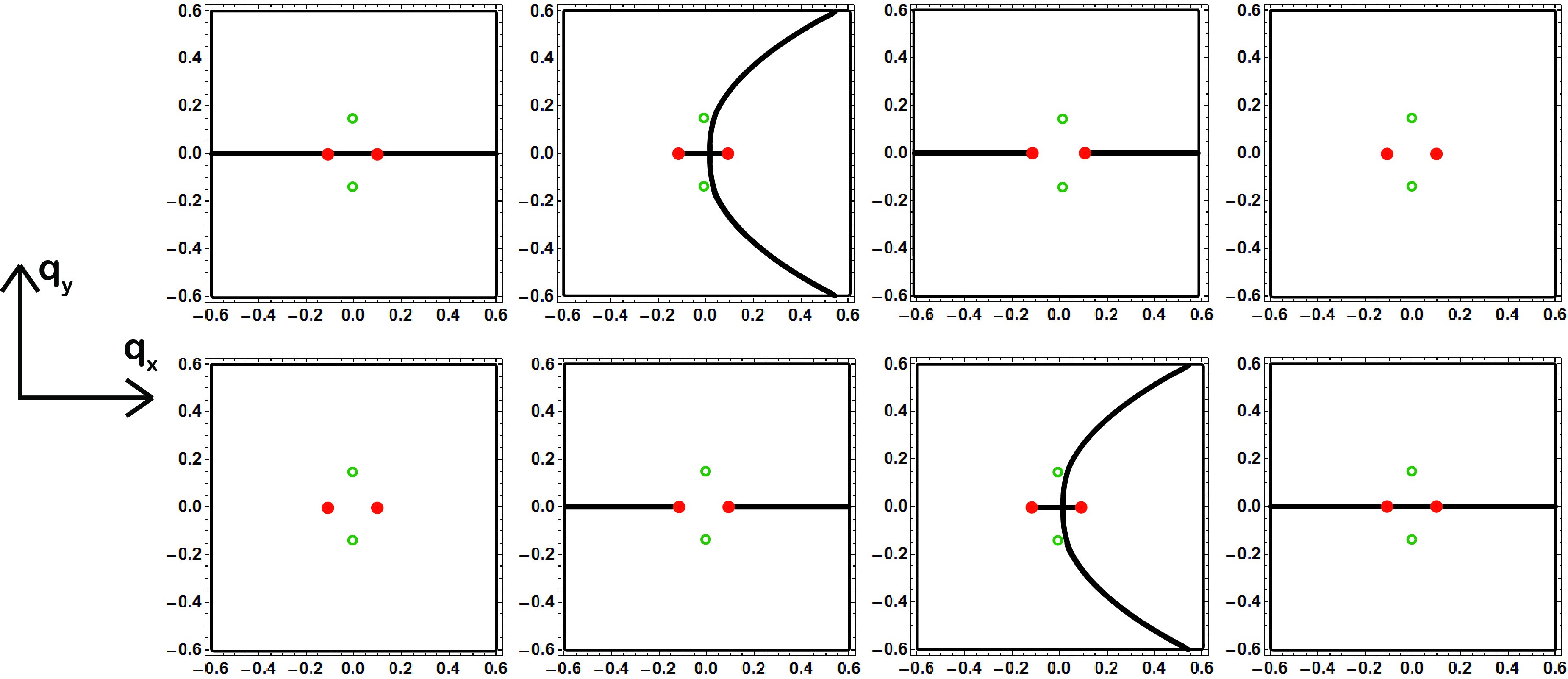}
  \caption{\label{Model2supNodes} Nodal lines in the photoemission amplitudes calculated for interlayer coupling Model II with $\hat V_{\rm int} \propto  \sigma_x$
  in the strong coupling regime. The top row is for $x$-polarized light and the bottom row is for $y$ polarized light. The four columns are for the four band
  of the model ordered by energy. The data show the effect of reconstruction of
  the $E=0$  Dirac points (red dots) in the strong coupling regime which are positioned along the $\pm q_x$ axis with the
  geometrically defined Dirac points (open green circles) along the $\pm q_y$ axis.  The nodal lines in bands 2 and 3 terminate on these renormalized Dirac points. An
  additional nodal line along the midline occurs due to interlayer interference of the transition amplitudes. }
\end{center}
\end{figure}

\end{widetext}

In the strong coupling regime the nodal patterns in Model II shown in Figure \ref{Model2supNodes} are more interesting. Here
the reconstructed strong coupling Dirac points occur along the high symmetry $q_x$ axis. Thus the interlayer nodal lines can
and do terminate at singular points along the $q_y$ axis. Interestingly these spectra also support branching points as shown.
The nodal lines that branch away from these branch points propagate without termination to large momenta as shown.

\subsubsection{Model III}

In Model III the interlayer coupling matrix $\hat V_{\rm int} \propto \mathbb{I}$ so that the coupling is dominated by
amplitudes that connect the same sublattice in the two layers. The nodal patterns calculated for this model are shown in Figure
\ref{Model1Nodes}. The signature of this family of interaction models is the appearance of second generation symmetry-protected
Dirac points at finite energy.

\begin{widetext}

\begin{figure}
\begin{center}
  \includegraphics[angle=0,width=0.8\columnwidth]{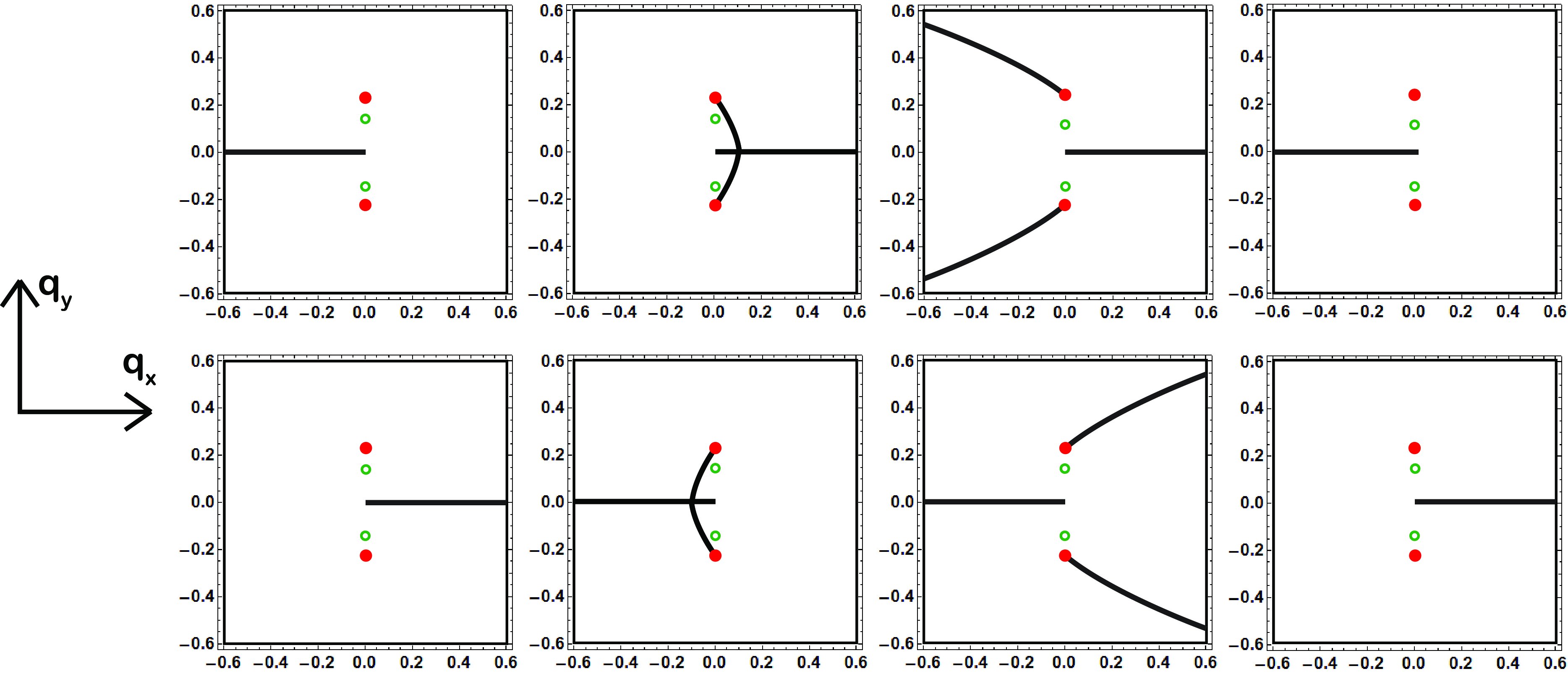}
  \caption{\label{Model1Nodes} Nodal lines in the photoemission amplitudes calculated for interlayer coupling Model III with $\hat V_{\rm int} \propto  \mathbb{I}$.
  The top row is for $x$-polarized light and the bottom row is for $y$ polarized light. The four columns are for the four bands
  of the model ordered by energy. Red dots denote the renormalized $E=0$ Dirac points in the model that are
  displaced away from each other along the $\pm q_y$ axis by gauge potentials generated by the interlayer coupling. The open green
  circles denote the undisplaced Dirac nodes of two decoupled layers.   The nodal lines in bands 2 and 3 terminate on these renormalized Dirac points. An
  additional one-sided nodal line  terminating at a second generation Dirac point occurs along the midline occurs in all four bands due to interlayer interference of the transition amplitudes.}
\end{center}
\end{figure}

\end{widetext}

Interlayer nodal lines propagating along the $q_x$ axis collide with these second generation points where they switch between
bands. The fundamental signature of this class of models is the appearance of new one-sided nodal lines along the midline
$q_y=0$. This behavior is clearly evident in both polarizations in Figure \ref{Model1Nodes}. The nodal lines that terminate on
the renormalized $E=0$ Dirac points are again curves rather than straight rays, manifesting the $\bm{q}$-dependence of the
 gauge
potential generated by the interlayer coupling.

\subsubsection{Phases along the midline}

Figure \ref{alphaplot} shows the internal phases $\alpha$ in the wavefunctions $\Psi_\pm$ calculated along the midline $q_y=0$
using Eqn. (\ref{alpha}) for the three interaction models discussed in the previous sections. We have verified that the the
internal phase $\alpha$ determined by differentiation of the emission intensities is independent of the value of the phase lag
$\phi$ for emission from the two layers. A careful inspection of these plots allows one to further discriminate between the
various interaction models.

All the models are characterized by an evolution from $\alpha = 0$ to $\alpha = \pi/2$ as a function of increasing $q_x$. The
phase angle $\alpha$ is {\it half} the phase difference between sublattice amplitudes in the same layer. Thus this evolution
describes the transition from states composed of the asymptotic (large $|q_x|$) layer eigenstates $(1,1)$ to $(1,-1)$ which is
a common feature in the negative energy solutions in all three interaction models.

The intermediate behavior is quite different in these models and can be used to analyze the symmetry of the interlayer coupling
in the intermediate and small $q_x$ region. For example, in Model II (middle panel) the interlayer coupling matrix $\hat V_{\rm
int} \propto \sigma_x$ and using Eqn. (\ref{topbottom}) the phase $\alpha$ is then identified as the geometrical Dirac angle
$\arctan(\Delta K/q_x)$. Note that this is the {\it same} for both negative energy bands at all values of $q_x$ in this model.
By contrast Model III (right panel) shows a different phase angle $\alpha$ for the two negative energy states as a consequence
of their interlayer coupling. The signature of this family of interaction models is both this phase splitting and a jump
discontinuity that occurs at $q_x=0$ where the two negative energy bands touch and the $R$-parity of a given band changes.
Model I combines features of both: the phase angle $\alpha$ deviates from the Dirac angle defined by the rotational
misorientation and the angles $\alpha$ evolve smoothly from $0$ to $\pi/2$ in both manifolds.

\begin{widetext}

\begin{figure}
\begin{center}
  \includegraphics[angle=0,width=0.8\columnwidth]{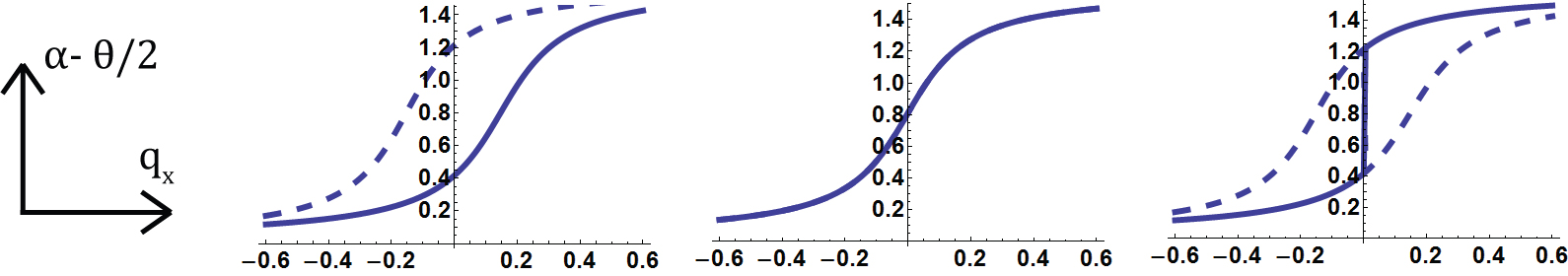}
  \caption{\label{alphaplot} Variation of the internal phase parameter $\alpha$ as a function of $q_x$ along the midline
  $q_y=0$. The three panels are for three limiting coupling models: left panel: $\hat V_{\rm int} \propto {\mathbb I}
  +\sigma_x$, middle panel: $\hat V_{\rm int} \propto \sigma_x$, right panel: $\hat V_{\rm int} \propto {\mathbb I}$. The solid(dashed)
  curves give the variation of the phase angle $\alpha$ in the first(second) occupied band. }
\end{center}
\end{figure}

\end{widetext}

\section{Discussion}

In this paper we have seen how the energy and polarization dependence of ARPES momentum distributions from bilayer graphene can
be used to measure the phase structure of its layer-coupled wavefunctions. The nodal surfaces discussed in this paper can be
measured in two ways. One may sweep the excitation energy and keep the outgoing electron energy fixed, in which case one
measures the momentum distribution as a function of a varying initial state energy. This protocol effectively sweeps the
constant initial state energy contours through a pattern of nodal lines in their transition probabilities and allows a
determination of momentum space nodal structures shown in Figures 7-10. The evolution of the momentum distributions during this
type of sweep are illustrated in Figure \ref{stacked} for interaction Models II and III. Alternatively, one can sweep the
excitation energy and detection energy simultaneously, thereby measuring the momentum distribution for a single initial state
energy. This protocol probes the symmetry of single initial state wavefunctions. Both methods are useful. The fingerprint of
intralayer interference are nodal patterns which do not change as a function of the excitation energy, and this seems best
suited to the first method. By contrast, interlayer interference effects {\it are} energy dependent and our analysis exploits
the differential energy dependence of the emission intensities in orthogonal polarizations. This requires following the
emission from the {\it same} intial state as the excitation energy is varied and can be analyzed most easily by the second
method.

It will be useful to first confirm these predicted energy dependences experimentally. We expect there will be a clean
separation between intra- and inter- layer interference effects seen in the experimental spectra with linearly polarizated
light and a suppression of these effects for circularly polarized light. Then focusing on the linearly polarized spectra,
observation of the curvature of the intra-layer nodal surfaces that terminate on the primary Dirac points will provide a
definitive experimental signature of the momentum dependence of the gauge potentials produced by the interlayer coupling.
Conversely, observation of $\phi$-periodic variations of the emission intensities along particular lines in momentum space
identify those features that are controlled by the interlayer coherence of the electronic states and  differential dependence
of their intensities on the excitation energy can be used to determined the sublattice symmetry of the matrix valued interlayer
coupling potential.

Although we have focused on the situation with no vertical electrostatic potential difference between the layers  the results
can be generalized to graphene bilayers with scalar layer asymmetry. In this case the relevant momentum space contours deform
to a family of mutually orthogonal confocal hyperbolas and ellipses. The extension of our methods to that situation is
relatively straightforward and will likely be needed for a quantitative analysis of experimental data. The methods developed in
this paper are also quite general and can be implemented with more sophisticated models for the electronic structure of BLG. It
is hoped that measurement and analysis along these lines will provide a clean experimental resolution of the nature of the
electronic states in twisted
bilayer graphenes.

\begin{widetext}

\begin{figure}
\begin{center}
  \includegraphics[angle=0,width=0.8\columnwidth]{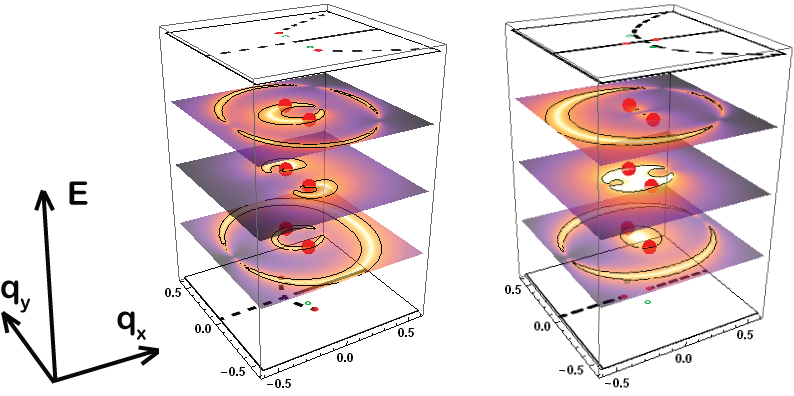}
  \caption{\label{stacked} Illustration of the evolution of ARPES momentum distributions for three initial energies crossing through the
  nodal lines for two different interaction models. (Left panel) Model III with $\hat V_{\rm int} =0.15 {\mathbb I}$. (Right panel) Model II
  in its strong coupling regime with $\hat V_{\rm int} = 0.2 \sigma_x$. The floor(ceiling) in these plots show the nodal
  surfaces superposed from the negative(positive) energy bands respectively.}
\end{center}
\end{figure}

\end{widetext}

\section{Acknowledgements}
This work was supported by the Department of Energy, Office of Basic Energy Sciences under Grant No. DE-FG02-ER45118.

\end{document}